\documentclass[epsf,amsfonts,amssymb,12pt]{article}
\usepackage{amsmath,amsfonts,amssymb,booktabs,multirow,cite}
\usepackage{array}
\usepackage{graphicx}
\usepackage{xcolor}
\usepackage{float}
\usepackage{placeins}
\usepackage[unicode]{hyperref}
\hypersetup{
colorlinks=false,
linkbordercolor=blue,
citebordercolor=red,
urlbordercolor=cyan
}
\usepackage{subcaption} % or subfigure (older)
\usepackage{comment}  
\usepackage[switch]{lineno}
\newif\iffastcompile
\fastcompiletrue

\usepackage{ascmac} % itemboxに要
\usepackage{amssymb,amsmath,amsfonts,mathrsfs} 
\usepackage{tikz}

\usetikzlibrary{calc}
\usetikzlibrary{decorations.markings,decorations.pathmorphing}

\usepackage{amsmath,amsfonts,latexsym,amssymb}
\usepackage{mathrsfs}
\usepackage{xcolor}
\usepackage{bm}
\usepackage[normalem]{ulem}

\definecolor{DarkBlue}{rgb}{0,0,0.9}

\definecolor{DarkRed}{rgb}{0.65,0,0}

\definecolor{DarkGreen}{rgb}{0.0,0.6,0.0} 
\definecolor{RefPurple}{rgb}{0.45,0.0,0.65}

\def\bena{\begin{eqnarray}}
\def\eena{\end{eqnarray}}
\newcommand{\non}{\nonumber}

\makeatletter

\def\affiliation#1{\def\@affiliation{#1}}

\def\@maketitle{
\begin{center}
{\Large \bf \@title \par}
\vspace{5mm}
{\normalsize \@author \par}
{\normalsize \it \@affiliation \par}
\end{center}
}

\@addtoreset{equation}{section}

\makeatother

\title{
\begin{flushright}
\end{flushright}
\vspace{0.5em}
\hfill\normalsize\textbf{Report number: NU-QG-15}\\[0.8em]
\Large Ringdown waves from hairy black holes}

\medskip 

\author{
Ariadna Uxue Palomino Ylla${}^{a}$\footnote {palomino.ylla.ariadna.uxue.e8@s.mail.nagoya-u.ac.jp}, ~
Kosuke Makino${}^{b}$%
, ~Akane Tanaka${}^{b}$%
, \\
~Akihiro Ishibashi${}^{a,c}$%
, and 
Chul-Moon Yoo${}^{a,c}$
}

{}\bigskip 

\affiliation{$^a$
Graduate School of Science,
Nagoya University, Nagoya 464-8602, Japan
\\
${}^{b}$Department of Physics, 
Kindai University, Higashi-Osaka, Osaka 577-8502, Japan
\\
$^c$
Kobayashi-Maskawa Institute for the Origin of Particles and the Universe (KMI), \\
Nagoya University, Nagoya 464-8602, Japan
} 

\medskip 

\date{}

\begin{document}
\iffastcompile\else\linenumbers\fi
\maketitle

\begin{abstract}
We study how quasinormal-mode frequencies may encode information about the effective matter source responsible for black-hole hair. Using the established
eikonal correspondence between quasinormal modes and unstable null geodesics, we relate shifts in the ringdown spectrum to perturbations of the photon-orbit frequency and Lyapunov exponent.
The black-hole hair is treated as an anisotropic fluid perturbatively added to vacuum black holes (Schwarzschild and Kerr black holes). In particular, we derive formulas that allow one to directly read off deviations from the Schwarzschild or Kerr QNM spectrum in terms of the corresponding equation-of-state parameters of the anisotropic fluid. 
In this setting, independent of energy conditions, our formulas offer a systematic method to compute quasinormal-mode frequencies for a broad class of hairy black holes. 
\end{abstract}

\section{Introduction}

Recent cosmological and astrophysical observations have accumulated compelling evidence for the existence of black holes, making them prime candidates for testing general relativity. On the theoretical side, a variety of non-vacuum or hairy black hole models have been proposed. These models may be interpreted as effective descriptions of black holes surrounded by dark-sector fields or the effects of modified-gravity theories near a black hole. It is highly relevant to investigate the basic characteristics of hairy black holes and to develop techniques for identifying them, for instance, through gravitational wave measurements. When a black hole is perturbed, the remnant emits gravitational radiation known as the ringdown phase. This signal is primarily characterised by a collection of damped oscillations whose complex frequencies correspond to the quasinormal modes (QNMs). These modes encode essential information about the black hole’s physical properties. If QNMs from a black hole endowed with “dark hair” could be detected, they would allow us to determine not only the mass and angular momentum of the source but also the state parameters of the surrounding dark-sector fields or the coupling constants that define the underlying modified gravity theory.

\medskip 

However, apart from a few particular cases, such as the analysis of QNMs from the Bardeen black hole by Konoplya et al.~\cite{Konoplya:2023ahd}, a systematic study of QNMs applicable to a broad class of hairy black holes has not yet been fully pursued. This is in part because the metrics describing hairy black hole models are generally complicated, making direct QNM calculations challenging.
Standard QNM methods require deriving the relevant perturbation equation for each background and perturbing field, imposing QNM boundary conditions, and solving for the complex frequencies using, for example, WKB methods, continued-fraction techniques, or time-domain evolution~\cite{Berti:2009kk,Iyer:1986np,Iyer:1986nq}. These methods are the most appropriate tools for obtaining accurate, field-dependent spectra, especially for the low-$\ell$ modes relevant to gravitational-wave observations, but they require the effective wave potential to be known for each model and perturbation sector.

\medskip

A closely related approach is the parametrized or perturbative ringdown framework, where small deformations of the effective wave potential are mapped to shifts of the QNM frequencies~\cite{Cardoso:2019mqo,McManus:2019ulj}. This provides a systematic way to study deviations from known black-hole spectra, including field-dependent effects and coupled perturbation equations. However, since the deformation is introduced at the level of the wave potential, the relation between the resulting QNM shift and the physical source supporting the geometry is not always direct. This motivates the development of complementary analytic formulas that can be applied uniformly to the leading QNM shifts of different hairy black-hole models, independently of the detailed matter model or dark-sector equation of state.

\medskip

In this paper, we consider hairy black hole models constructed by perturbatively adding anisotropic fluids, mimicking dark sector fields, to vacuum black hole metrics, namely the Schwarzschild and Kerr metrics. To systematically understand how QNMs of these constructed hairy black holes are modified by their non-trivial hair, compared to their corresponding vacuum counterparts, we exploit the eikonal/WKB correspondence between unstable circular orbits of photons (UCOPs) and QNMs.
In this correspondence, the real part of the QNM frequency is related to the orbital frequency $\Omega$ of the unstable null orbit, while the imaginary part is controlled by the corresponding Lyapunov exponent $\lambda$. For static spherically symmetric spacetimes this gives
\bena
\omega_{\rm QNM} = \Omega \ell -i(n+1/2) \lambda \,,
\label{eq:QNMStatic}
\eena
where $\ell$ is the angular quantum number and $n$ is the overtone index. The QNMs are also labelled by the azimuthal quantum number $\mathbf{m}$, whose sign is degenerate in the non-rotating vacuum case.

\medskip

Historically, the QNM--geodesic correspondence can be traced back to the geometrical interpretation of black-hole ringing as waves temporarily trapped near unstable null orbits, first discussed by Goebel~\cite{Goebel:1972}. It was subsequently developed into the WKB/geodesic picture of black-hole quasinormal modes~\cite{Ferrari:1984zz,Mashhoon:1985cya,Schutz:1985km,Iyer:1986np,Cardoso:2008bp}. Further justification using the Penrose limit was discussed in Refs.~\cite{Fransen:2023eqj,Kapec:2024lnr,Yoo:2025tzq}. In the present work, we use this correspondence as an established tool rather than as a result to be proven. Our contribution is to combine it with the Einstein equations to express the leading eikonal shifts directly in terms of the effective matter variables of the hairy black-hole geometry.

\medskip

The main distinction from potential-based parametrisations is therefore the source-level interpretation. Once the metric, or equivalently the effective stress-energy tensor, is specified, the same formulas give the leading shifts of the photon-orbit frequency and Lyapunov exponent. In the static case, for instance, the difference between the fractional damping and oscillation shifts is controlled by the local combination $\rho+P_\theta$ at the photon orbit. Thus, the sign and magnitude of the relative damping shift can be interpreted in terms of the tangential null-energy combination of the effective source. This dependence on $\rho+P_\theta$ is not manifest in a generic deformation of the effective wave potential.

\medskip

We emphasise that the QNM--UCOP relation used in this work is an eikonal/WKB correspondence rather than an exact statement about the full quasinormal-mode spectrum. In this regime, the angular quantum number is large, and the overtone number is kept comparatively small, namely $\ell \gg 1$ and $n \ll \ell$. Therefore, the approximation is not expected to provide precise values for the dominant gravitational-wave ringdown mode, which is usually the fundamental $(\ell,\mathbf{m},n)=(2,2,0)$ mode~\cite{Giesler:2019uxc}. As a simple benchmark, in Schwarzschild spacetime the leading geodesic estimate gives
\begin{equation}
M\omega_{\rm eik}\simeq \frac{2}{3\sqrt{3}}-\frac{i}{6\sqrt{3}}
\simeq 0.3849-0.0962i \, ,
\end{equation}
for $\ell=2$ and $n=0$, while the accurate gravitational value is $M\omega\simeq0.3737-0.0890i$~\cite{Berti:2009kk}. Thus, even for the dominant mode, the leading eikonal estimate captures the oscillation frequency and damping rate at the level of a few to ten percent in the Schwarzschild case, while its accuracy improves in the true eikonal regime.

\medskip

The applicability of the correspondence also requires a well-behaved WKB effective potential, with a single dominant maximum, two turning points, and the appropriate decay toward the relevant boundaries, such as the horizon and spatial infinity, or the outer boundary in non-asymptotically flat cases. In the present work, we consider black-hole-type geometries obtained as small deformations of Schwarzschild or Kerr. In this perturbative regime, the unstable circular photon orbit and the associated WKB peak are expected to be smooth deformations of their vacuum counterparts. The explicit examples studied below, including the Bardeen, Hayward, and Kiselev metrics, illustrate that the geodesic construction can be applied systematically in this regime.

\medskip

A second, independent restriction is that the QNM--UCOP correspondence is most directly guaranteed for test fields propagating on a fixed black-hole background, and not in general for gravitational perturbations themselves or for fields non-minimally coupled to gravity~\cite{Konoplya:2017wot,Konoplya:2022gjp}. Our calculation satisfies the fixed-background/geodesic part: we treat the hairy metric as a fixed geometry and compute the first-order shifts of its unstable circular null orbit, orbital frequency, and Lyapunov exponent. When these shifts are interpreted as gravitational ringdown/QNM shifts, we are therefore making the additional eikonal assumption that the relevant gravitational perturbation sector follows the same photon-orbit correspondence. Thus, our formulas should be understood as first-order leading-eikonal predictions for the expected shifts in frequency and damping rate, rather than as a full calculation of the gravitational QNM spectrum for each hairy black-hole model.

\medskip

By expressing the eikonal QNM corrections in terms of anisotropic-fluid variables, we provide a framework in
which deviations from the Schwarzschild or Kerr QNM spectrum can be interpreted as constraints on the associated equation-of-state parameters.
We then express $\Omega$ and $\lambda$ explicitly in terms of the state parameter of the dark sector field. In this context, it is worth noting the no-short hair theorem, according to which, if a static black hole supports hair in the form of an anisotropic fluid satisfying certain energy conditions, that hair must extend beyond the radius of the UCOP, namely the photon sphere~\cite{Ishibashi:2023vsm}.   

\medskip

We also perform a similar analysis for stationary rotating hairy black holes. In rotating spacetimes, the QNM--photon-orbit correspondence is not justified in the same general form as in the static, spherically symmetric case. Rotation breaks the $\mathbf{m}$-degeneracy and therefore, generic eikonal QNMs are associated with spherical photon orbits rather than only with equatorial circular photon orbits. In this work, we restrict the rotating analysis to orbital rays trapped in the equatorial plane. This corresponds to the eikonal sector with $\ell=|\mathbf{m}|$, describing the co-rotating and counter-rotating equatorial modes~\cite{Berti:2005eb,Berti:2005ys,Yang:2012he}. In that restricted sector, the rotating analysis is analogous to the static case, because the relevant QNM shifts are estimated from the shift of the equatorial photon-orbit frequency and Lyapunov exponent. However, this should not be interpreted as a description of the full rotating QNM spectrum. Related developments include the QNM-shadow correspondence and its applications to modified, quintessence-like, and rotating black holes~\cite{Stefanov:2010xz,Jusufi:2020dhz,CuadrosMelgar:2020kqn,Jusufi:2022,PedrottiVagnozzi:2024,YuChenGao:2022}.

\medskip

Recently, Igata developed a coordinate-invariant formulation relating strong-deflection-limit coefficients, local curvature, matter variables, and eikonal QNMs~\cite{Igata:2025taz,Igata:2025plb,Igata:2025hpy}. In the static spherically symmetric case, the logarithmic-divergence coefficient can be written directly in terms of the local combination of energy density and tangential pressure at the photon sphere. This is closely related to our use of the local matter combination $\rho+P_\theta$ in the eikonal damping shift.

\medskip

The paper is organised as follows: In the next section, we derive the UCOP for a general static spherically symmetric black hole geometry. We also discuss the relation between the UCOP and QNMs. In section \ref{ch:examplesStatic},  as our examples of hairy static black holes, we examine the cases of Bardeen, Hayward, and Kiselev metrics, to demonstrate how our method works. In section~\ref{ch:rotating}, we derive general formulas for stationary rotating hairy black holes.
In section \ref{ch:examplesRotating}, we examine the rotating version of the models previously introduced in section \ref{ch:examplesStatic}.
In the appendix, we describe some useful geometric formulas for our analyses.
This paper uses the geometrized units convention ($G =1$, $c=1$).

\section{Static hairy black holes and QNMs}\label{ch:Static}

In this section, we evaluate the components ($\Omega$, $\lambda$) of the QNMs~\eqref{eq:QNMStatic} for static spherical hairy black holes from the orbital frequency and Lyapunov exponent for the UCOP. 

\subsection{Static hairy black holes and null geodesics} \label{subsec:static:formulas}

We start with the following general static spherically symmetric metric 
\bena
 ds^2 = - f(r) dt^2 + h(r) dr^2 + r^2 (d\theta^2 + \sin^2 \theta d \varphi^2) 
 \,. 
 \label{metric:static:gen}
\eena
To consider geodesic curves on this background, let ${\cal L}$ be the Lagrangian, 
\bena
 {\cal L} := \dfrac{1}{2}\left( -f \dot{t}^2 + h \dot{r}^2 + r^2 \dot{\varphi}^2 \right) = \dfrac{\epsilon}{2} \,,
\label{eq:L1}
\eena  
where an overdot denotes differentiation with respect to the affine parameter $\sigma$ along the geodesic, e.g. $\dot{x}^\mu \equiv dx^\mu/d\sigma$. For timelike geodesics, $\sigma$ may be chosen as the proper time. We have already set $\theta = \pi/2$ and $\epsilon = -1$ for timelike geodesics and $\epsilon =0$ for null geodesics. 
From the isometries along $t$ and $\varphi$ we immediately find the two conserved quantities
\bena
 E:= - \dfrac{\partial {\cal L}}{\partial \dot{t}} = f\dot{t} \,, \quad 
 L:= \dfrac{\partial {\cal L}}{\partial \dot{\varphi}} = r^2 \dot{\varphi} \,.
\eena
Plugging these two into the Lagrangian~\eqref{eq:L1}, we obtain
\bena
  \dfrac{1}{2}\dot{r}^2 + V(r) = 0 \,, \quad V(r) := \dfrac{1}{2\, h(r)}\left(-\epsilon + \dfrac{L^2}{r^2}-\dfrac{E^2}{f(r)} \right) \,. 
\eena

For convenience we define $A(r):= -\epsilon + \frac{L^2}{r^2} - \frac{E^2}{f}$. Therefore $V(r) = \frac{A(r)}{2\, h(r)}$.
The conditions for geodesic curves to be spatially closed are given by 
\bena
 V=0 \,, \quad V' = 0 \,, 
 \label{eq:zerogeodesics}
\eena
on the orbit, where the prime denotes the derivative by $r$. 
From the above two conditions, we obtain, respectively, 
\bena
A(r)\circeq 0 \, \Rightarrow \,\dfrac{L^2}{r^2}=\dfrac{E^2}{f}+\epsilon \,,
\label{eq:zeroes0}\\[-0.2em]
\frac{A'(r)}{2\,h(r)}-\frac{h'(r)}{2 \, h(r)^2}A(r) \circeq 0\, \Rightarrow \, \dfrac{2L^2}{r^3} = \dfrac{f'}{f^2}E^2 \, ,
\label{eq:zeroes1}
%\dfrac{2L^2}{r^3} = \dfrac{f'}{f^2}E^2 
%\Leftrightarrow
%-\frac{h'(r)}{h(r)} V(r) + \frac{1}{2 h(r)}\left(-2\frac{L^2}{r^3}+\frac{f'(r)E^2}{f(r)^2}\right)\circeq 0 \,
\eena

where here and hereafter ``$\circeq$" implies the equality holds on the circular orbit under consideration. The term proportional to the first derivative of $h(r)$ in terms of $r$ in \eqref{eq:zeroes1} disappears because it is proportional to $A(r)$ and given \eqref{eq:zeroes0}, when evaluated at the photon orbit, it would become $0$. Combining these two equations, we obtain 
\bena
 \epsilon \circeq \dfrac{E^2}{2f^2}(rf'-2f) \,, 
\label{epsilon}
\eena

Using $V=0=V'$, we can write $V''$ as 
\bena
 V'' \circeq \dfrac{1}{2h}\left[ 6\dfrac{L^2}{r^4}+\left( \dfrac{f'}{f^2}\right)' E^2 \right] \,. 
\label{eq:V''} 
\eena
In the expression of $V''$, the terms associated with derivatives of $h(r)$ become zero, because they are also multiplied by $A(r)$ or $A'(r)$, which are zero when evaluated at the geodesic according to \eqref{eq:zeroes0} and \eqref{eq:zeroes1}.

Let us focus on null geodesic curves describing UCOP. Combining Eqs.~(\ref{eq:zeroes1}) and \eqref{eq:V''} with $\epsilon =0$, we have 
\bena
 V'' \circeq \dfrac{L^2}{2r^2h}\left( \dfrac{f''}{f} -\dfrac{2}{r^2} \right) \,. 
\eena

Then, following the reference~\cite{Cardoso:2008bp}, we obtain the Lyapunov exponent $\lambda$ and the orbital frequency $\Omega$ as 
\bena
 \lambda^2 := -\dfrac{V''}{\dot{t}^2} &\circeq & \dfrac{f}{2h}\left( \dfrac{2}{r^2}-\dfrac{f''}{f} \right) \,, 
 \\
 \Omega := \dfrac{\dot{\varphi}}{\dot{t}} & \circeq& \dfrac{\sqrt{f}}{r} \,.  
\label{eq:orbitals1}
\eena
For the Schwarzschild metric case, i.e., $f=h^{-1}=1-2M/r$, we obtain the well-known result
\bena
 \lambda_0= \dfrac{1}{3\sqrt{3}M} \,, \quad \Omega_0 = \dfrac{1}{3\sqrt{3}M} \,. 
\eena

\subsection{Anisotropic fluid surrounding black holes and QNM frequency}

Let us consider the case in which our metric solves the Einstein equations. 
%As the matter fields mimic dark sector fields, w
Here, we construct the background metric assuming the leading order metric is given by the Schwarzschild metric, namely, 
$f(r)=f_0(r)+\mathcal O(\eta)$ and $h(r)^{-1}=f_0(r)+\mathcal O(\eta)$ with $f_0(r)=1-2M/r$ and $\eta$ being a dimensionless small parameter. 
At the next-leading order, we consider fluids with anisotropic pressure so that the stress-energy tensor $T^\mu{}_{\nu}$ is given 
by the components, 
\bena
 T^t{}_t= - \rho \,, \quad T^r{}_r = P_r \,, \quad T^\theta{}_\theta = T^\varphi{}_\varphi = P_\theta \,, 
\label{eq:stressenergy}
\eena
and the rest of the components are vanishing. Combining the components of the Einstein tensor, $G^0_{0} = -8\pi  \rho$, $G^r{}_r= 8\pi P_r$, and $G^\theta{}_\theta=8\pi P_\theta$, we obtain
\bena
 \lambda^2  \circeq \dfrac{f}{r^2}
           -\dfrac{f}{r^2}\left[ 2\pi r^2(\rho-3P_r +4P_\theta) + 2\pi r^2 h(\rho+P_r) + 16\pi^2 r^4 hP_r(\rho+P_r)\right] \,. 
\label{eq:lambda:gen}
\eena 
Now, let us suppose that the equations of state for our fluid are given by two parameters  
\bena
 P_r =w_r \rho \,, \quad P_\theta = w_\theta \rho \,.
 \label{eq:ws}
\eena

 Since the matter variables are treated as first-order perturbations,
\bena
\rho,P_r,P_\theta=O(\eta)\,,
\eena

with $0<\eta\ll1$. 
More specifically, we assume
\bena
\left|4\pi\int^r_{r_0}\rho r^2 dr\right|/M\sim \mathcal O(\eta) \ll 1.  
\eena

\medskip 

To obtain fully linearized expressions for the QNM coefficients around the Schwarzschild background, we further need to expand 
\begin{align}
&r_\star      \simeq r_0 + \delta r = 3M + \delta r \,,\\
&H(r) \equiv h(r)^{-1} \simeq \, f_0 (r) + \delta H(r)\,, \\ 
&f(r)         \simeq f_0(r) + \delta f(r) \,,\\
&\lambda_\star\simeq \lambda_0 + \delta \lambda = (3\sqrt{3}M)^{-1} + \delta \lambda \,,\\
&\Omega_\star \simeq \Omega_0 + \delta \Omega = (3\sqrt{3}M)^{-1} + \delta \Omega \,.
\end{align}
where $r_0 = 3M$ denotes the Schwarzschild UCOP radius. 
The terms $\delta r$, $\delta H$ and $\delta f$ represent small linear deviations of $\mathcal O(\eta)$ from these Schwarzschild quantities
sourced by the anisotropic fluid. 
% \begin{equation}
%   \delta f, \delta H = O(\eta)\,.
% \end{equation}

Let $r_\star$ denote the UCOP radius of the hairy black hole. Throughout the following, a subscript $\star$ indicates evaluation at this radius.
Since the following terms are already proportional to the
matter variables, products such as \(\delta h\,\rho\), \(\delta h\,P_r\), or \(\delta f\,\rho\) are of order \(O(\eta^2)\) and are neglected in the present first-order treatment. Thus, the leading-order expression for $\lambda$~\eqref{eq:lambda:gen} evaluated at the UCOP becomes
\bena
 \lambda_\star &\simeq& \Omega_\star - \Omega_\star \pi r_\star^2 \rho \left[ 1-3 \, w_r +4 \, w_\theta + \dfrac{(1+w_r)}{f_0} \right]\bigg|_{r=r_\star}
 \label{lamda}\,,  
\eena
where $\Omega_\star = \dfrac{\sqrt{f(r_\star)}}{r_\star}$ represents the orbital frequency at the UCOP. 

We start by substituting values in the UCOP expression $g(r)= r f'(r) - 2 f(r)$, derived from~\eqref{epsilon}, which vanishes for the circular photon orbit. By evaluating it in the UCOP radius for the hairy black hole, we obtain
\bena
g(r_\star) \simeq ((r_0 + \delta r)(f_0'(r) + \delta f'(r)) - 2 (f_0(r_0 + \delta r) + \delta f(r_0 + \delta r)))|_{r =  r_0 + \delta r} \,.
\eena
%Expanding around $r_0$ and keeping terms up to the first order we obtain
%\bena
%g(r_*) \simeq (r_0 + \delta r)(f_0'(r_0) + f_0''(r_0) \,\delta r + \delta f'(r_0)) - 2 (f_0(r_0) + f_0'(r_0)\,\delta r + \delta f(r_0)) \,.
%\eena
By using the result for the Schwarzschild photon sphere $g(r_0) = r_0 f_0'(r_0) - 2 f_0(r_0) = 0$, and by keeping perturbation terms up to the first order, we can solve this expression for $\delta r$ as
%\bena
%g(r_*) \simeq - \delta r (f_0'(r_0) - r_0 \, f_0''(r_0)) + r_0  \delta f'(r_0) - 2\, \delta f (r_0) = 0\,.
%\eena
%Solving for $\delta r$, we obtain 
\bena
\delta r \simeq \frac{r_0 \,\delta f ' (r_0) - 2\, \delta f (r_0)}{f_0'(r_0) - r_0  f_0''(r_0)} \, = \,  \frac{3 M}{2} (r_0 \, \delta f ' (r_0) - 2 \, \delta f (r_0))\,.
\label{eq:deltar}
\eena

In a similar way, we proceed to linearise the orbital frequency~\eqref{eq:orbitals1} evaluated at the UCOP radius
we expand
\[
\Omega_\star^2
=
\frac{
f_0(r_0)+f_0'(r_0)\delta r+\delta f(r_0)
}{
(r_0+\delta r)^2
}
+O(\eta^2).
\]
By expanding the denominator, we obtain, up to first order
\[
\Omega_\star^2
\simeq
\frac{f_0(r_0)}{r_0^2}
+
\frac{\delta f(r_0)}{r_0^2}
+
\frac{\delta r}{r_0^2}
\left[
f_0'(r_0)-\frac{2f_0(r_0)}{r_0}
\right]\,.
\]
The term proportional to \(\delta r\) vanishes because it is proportional to the expression $g(r_0)=0$.
Therefore, the first-order displacement of the UCOP radius does not contribute
explicitly to \(\Omega_\star^2\), and we find

\bena
\Omega_\star^2
\simeq \frac{1}{3 r_0^2} (1 + 3 \, \delta f (r_0))\,,
\eena
which leads to
\bena
\Omega_\star
\simeq \Omega_0( 1 + \frac{3}{2} \delta f (r_0))\,.
\label{eq:lin_omega}
\eena

Finally, by substituting $\Omega_\star$  into the Lyapunov exponent (\ref{lamda}), we obtain
\bena
\lambda_\star 
\simeq \frac{1}{\sqrt{3} \, r_0} (1 + \frac{3}{2} \delta f) \, \bigg(1 - \pi r_\star^2 \rho (1 - 3 w_r + 4 w_\theta + \frac{(1 + w_r)}{f_0})\bigg)\bigg|_{r =  r_0} \,,
\eena
and by expanding up to the first order for the perturbations, we obtain 
\bena
\lambda_\star
\simeq \lambda_0 (1 - 4 \pi r_0^2 \rho (1 + w_\theta) + \frac{3}{2} \delta f)|_{r =  r_0 }\,.
\label{eq:lin_lamda}
\eena

\medskip 

\subsection{Modifications of QNMs and Energy Conditions}

In this section, we quantify the deviation of the QNMs for hairy black holes from their Schwarzschild counterparts, and relate them to the fluid state parameters $w_r$ and $w_\theta$. We can observe how it affects the QNM frequency as
\begin{equation}
\frac{\delta\Omega}{\Omega_0}=\frac{3}{2}\,\delta f(r_0),
\qquad
\frac{\delta\lambda}{\lambda_0}=\frac{3}{2}\,\delta f(r_0)-4\pi r_0^2 \rho\big[1+w_\theta\big]|_{r=r_0}.
\label{eq:OmegLambdaShifts}    
\end{equation}
Both shifts are controlled by the metric correction $\delta f(r_0)$ in the same way, but the Lyapunov exponent also presents an explicit contribution from the tangential pressure $P_\theta$. 

Although $\delta f$ appears geometric, it is not independent of the matter. To make this relation explicit, we now restrict the general metric~\eqref{metric:static:gen} to the subclass in which the radial metric component is the inverse of the temporal one, namely $h(r)=f(r)^{-1}$, or equivalently $H(r)\equiv h(r)^{-1}=f(r)$. This corresponds to setting the previously introduced perturbations consistently as $\delta H=\delta f$. Through the Einstein equations, this specialisation also fixes the radial equation-of-state parameter to $w_r=-1$, i.e., $P_r=-\rho$. This assumption is not overly restrictive for the examples considered below, since several standard regular or effective hairy black-hole models are commonly written in this form.
Then let us introduce the mass function $m(r)$ as follows:
\begin{equation}
f(r)=1-\frac{2m(r)}{r},\quad
h(r)=\left(1-\frac{2m(r)}{r}\right)^{-1},
\label{eq:metric-split}
\end{equation}
where $m(r) \simeq M+\delta m(r)$. Expanding to first order gives an explicit expression for the geometric term,
\begin{equation}
\delta f(r)=-\frac{2\,\delta m(r)}{r}.
\label{eq:deltaf}
\end{equation}

We now specify the convention used for the mass perturbation. The mass
parameter \(M\) is chosen to be the Schwarzschild mass appearing in the
asymptotic form of the hairy spacetime. Therefore,
\bena
m(\infty)=M,
\qquad
\delta m(\infty)=0 \,.
\eena
The perturbation \(\delta m(r)\) is not the accumulated mass from the
center up to \(r\). Instead, it measures the deviation of the local mass
function from its asymptotic Schwarzschild value. 
By using the Einstein equations (see Appendices), we can relate the metric function $m(r)$ to the matter field as 
\begin{equation}
m'(r)=\delta m'(r)=4\pi r^2\rho(r). %,\qquad
%\Phi'(r)=\frac{4\pi r^2\,(\rho(r)+P_r(r))}{\,r-2m(r)\,},
\label{eq:EinsteinSplit}
\end{equation}
By integrating this expression, we obtain 

% \[
% \delta m(\infty)-\delta m(r)
% =
% \int_r^\infty \delta m'(s)\,ds
% =
% 4\pi\int_r^\infty \rho(s)s^2\,ds .
% \]

\[
-\int_r^\infty \delta m'(s)\,ds
=
-\delta m(\infty)+\delta m(r)
=
\delta m(r)=-
4\pi\int_r^\infty \rho(s)s^2\,ds, 
\]
where we have used $\delta m(\infty)=0$. 
Thus, for \(\rho>0\), \(\delta m(r)<0\). This sign is a consequence of
setting the background Schwarzschild mass as its value at the asymptotic infinity. 

Through the Einstein equations, we can rewrite the QNM shifts directly in terms of the fluid parameters as follows
% and \(P_r\). 
\begin{equation}
\frac{\delta \Omega}{\Omega_0 } \simeq  \frac{12 \pi}{r} \int_r^{\infty} \rho(s)  s^2 \, ds\, |_{r =  r_0 },
\label{eq:eqORho}
\end{equation}
\begin{equation}
\frac{\delta\lambda}{\lambda_0} \simeq  \frac{12 \pi}{r} \int_r^{\infty} \rho(s)  s^2 \, ds - 4 \pi r_0^2 \rho (1 + w_\theta) |_{r =  r_0 }\, . 
\label{eq:eqLRho}
\end{equation}
As shown in Appendix~\ref{app4}, the conservation of 
% energy and momentum for 
an anisotropic fluid, expressed by the relation $\nabla_\mu T^{\mu}_{\nu} = 0$, leads to a generalized Tolman–Oppenheimer–Volkoff (TOV) equation. This relation connects the radial and tangential pressures as
\begin{equation}
P_\theta(r)
= P_r(r) + \frac{r}{2} P'_r(r) + \frac{(P_r+\rho)}{2}\left(\frac{4\pi r^3 P_r(r) + m(r)}{r-2m(r)} \right) \,,
\label{eq:TOVaniso}
\end{equation}
which shows that \(P_\theta\) 
% value 
is constrained once \(\rho\) and \(w_r\) are specified.

Energy conditions require normal and healthy properties of the matter field from a viewpoint of certain causality or stability. 
% restricts the behaviour of the matter field, in a way that it assures that the energy density is positive definite locally.
It is worthwhile to interpret the relationship between the properties of matter fields surrounding a black hole and QNMs in terms of energy conditions.
The most used conditions in general relativity are the Null, Weak, Strong, and Dominant energy conditions, which are locally defined and often abbreviated as: NEC, WEC, SEC, and DEC, respectively. WEC and SEC imply NEC, while DEC implies WEC. For their definitions and usages, see, e.g., a review \cite{Iizuka:2025xnd} and references therein.
Satisfying these conditions is necessary for interpreting the source as regular matter.
%, but it is not sufficient to guarantee physical viability.
For an anisotropic effective fluid with energy density $\rho$ and principal pressures $(P_r,P_\theta,P_\theta)$, the energy conditions then read
\begin{align}
\textbf{NEC:}&\quad \rho+P_r\ge 0,\ \ \rho+P_\theta\ge 0 
\  ,\label{eq:NEC}\\[4pt]
\textbf{WEC:}&\quad \rho+P_r\ge 0,\ \ \rho+P_\theta\ge 0 \ \&\ \ \rho\ge 0\
\ \Longleftrightarrow\ \rho\ge 0,\ w_r\ge -1,\ w_\theta\ge -1 ,\label{eq:WEC}\\[4pt]
\textbf{SEC:}&\quad \rho+P_r\ge 0,\ \ \rho+P_\theta\ge 0 \ \ \&\ \ \rho+P_r+2P_\theta\ge 0
\ ,\label{eq:SEC}\\[4pt]
\textbf{DEC:}&\quad |P_{r}|\le \rho , \ \, |P_{\theta}|\le \rho \, \Longleftrightarrow \, \rho\ge 0\ , \ \ |w_r|\le 1,\ |w_\theta|\le 1 \, .\label{eq:DEC}
\end{align}

In our barotropic model of an anisotropic fluid, with $i$ as an indicator of radial or tangential, the parameter $w_i =0$ corresponds to dust, $w_i=1/3$ to radiation, 
% $w_i<1/3$ 
$w_i<-1/3$
to dark-energy
% -like, 
% acceleration, 
and $w_i<-1$ to phantom energy. 
A particularly important case is $w_r=-1$ (cosmological-constant-like). Not only does it satisfy most energy conditions, but it is also the only scenario that ensures the continuity of the energy density across the black hole event horizon \cite{ChoKim2019} with non-zero energy density. In this scenario $\rho+P_r=0$ (at least on the horizon), and equation \eqref{eq:TOVaniso} reduces to
\begin{equation}
w_\theta = -1 - \frac{r}{2 } \frac{\rho'(r)}{\rho(r)} = - \frac{r}{2}\frac{m''(r)}{m'(r)}. 
\label{eq:wth1}
\end{equation}
By considering the expression~\eqref{eq:wth1}, we can derive an expression for the tangential pressure in terms of the second derivative of the mass function $P_{\theta}(r) = - \frac{m''}{8\pi r}$.
When considering $\rho(r)>0$, for regular matter, we find $\frac{\delta \Omega}{\Omega_0 } >0 $ from Eq.~\eqref{eq:eqORho}, which means the QNM of the hairy black hole oscillates faster than in the vacuum case. %Add a cite
In particular, it is also interesting to analyze
\begin{equation}
    \frac{\delta\lambda}{\lambda_0} - \frac{\delta\Omega}{\Omega_0} \simeq  - 4 \pi r_0^2 \rho (1 + w_\theta) |_{r =  r_0 }\,.
    \label{eq:dif1}
\end{equation}

Equation~\eqref{eq:dif1} is one of the central results of this section. It shows that, within the present perturbative setting, the difference between the fractional damping shift and the fractional oscillation shift is not controlled by the integrated mass correction, but by the local combination $\rho+P_\theta=\rho(1+w_\theta)$ at the photon orbit. Therefore, the sign of this difference gives a direct diagnosis of the tangential null-energy condition near the UCOP. If the effective matter has $\rho>0$ and satisfies the tangential NEC, then $\rho+P_\theta\ge0$ and the difference~\eqref{eq:dif1} is non-positive. Conversely, a positive value of $\delta\lambda/\lambda_0-\delta\Omega/\Omega_0$ would require $1+w_\theta<0$, and hence a violation of the tangential NEC in the effective matter description.

It is also interesting to examine the displacement of the UCOP radius position
\begin{equation}
    \delta r \simeq -12 \, \pi \, M \left(r_0^2 \, \rho(r) + \frac{3}{r_0} \int_r ^ \infty \rho(s) s^2 \, ds \right)\, ,
\end{equation}
which, for regular matter with positive energy density, is always negative. The UCOP radius position comes closer to the center when considering the presence of regular matter hair. 

% \section{Examples of Perturbative Hair Effects on QNMs}
\section{Examples of Static Spherical Hairy Black Holes}
\label{ch:examplesStatic}

In this section, we apply the formulas derived in the previous section to three specific static, spherically symmetric black hole models: the Bardeen, Hayward, and Kiselev spacetimes. The Bardeen and Hayward solutions are regular black holes that approach the Schwarzschild geometry at large distances. The Kiselev solution, in contrast, represents a non-vacuum black hole surrounded by an anisotropic fluid. Unlike the Bardeen and Hayward models, it typically features a central curvature singularity, and its asymptotic behaviour is determined by the chosen value of the equation-of-state parameter.

\subsection{Bardeen black hole} 

Let us consider the Bardeen black hole, whose metric is given by (\ref{metric:static:gen}) with the following components
\begin{equation}
 f(r)=1/h(r)= 1 - \dfrac{2Mr^2}{(r^2+q^2)^{3/2}} \,, 
\end{equation}
% with $q$ being a constant. 
where $q$ is a constant parameter. The Bardeen metric is interpreted as the solution of the Einstein equations with a certain type of nonlinear electromagnetic source (see e.g.,~\cite{Ayon-Beato:2000mjt,Rodrigues:2018bdc}), which allows the fluid expression as 
\begin{equation}
 \rho=-P_r=\dfrac{6M q^2}{8\pi(r^2+ q^2)^{5/2}} \,, \quad P_\theta=\dfrac{q^2 M (9r^2-6 q^2)}{8\pi (r^2+q^2)^{7/2}} \,.
 \label{eq:rho1}
\end{equation}
%We can read off $w_r=-1$, and hence 
%\bena
% \lambda \simeq \lambda_\star - 4\pi \lambda_\star (1+w_\theta)\rho \,. 
%\eena
We can read off the state parameters as
\begin{equation}
 w_r = -1 \,, \quad w_\theta =  \dfrac{3r^2-2 q^2}{2(q^2+r^2)}\,. 
\end{equation}

% Further, i
If we assume that the extra-parameter $q^2$ is sufficiently small, i.e., $|q/M|\ll 1$, we can treat the Bardeen solution as a perturbation from the Schwarzschild case and use the linearised expressions (\ref{eq:lin_lamda}) and (\ref{eq:lin_omega}). First, we estimate $\delta f(r)$ by expanding $f(r)$ up to first order
\begin{equation}
f(r) \simeq 1 - \frac{2M}{r} \left( 1 - \frac{3 q^2}{2 r^2} \right)\, \Longrightarrow \,\delta f (r) \simeq \frac{3 M q^2}{r^3}\, .
\end{equation}
%
% We calculate
Then we obtain
\begin{equation}
\Omega_\star \simeq \frac{1}{3\sqrt{3} M} \left( 1 + \frac{3}{2}\left(\frac{3 M q^2}{27 M^3}\right) \right) \,=\, \frac{1}{3\sqrt{3} M} \left(1+\frac{q^2}{6 M^2} \right)\,. 
\label{eq:O1}
\end{equation}
% and b
By keeping terms up to $q^2$, we approximately obtain $4\pi r_0^2 (\rho(r_0) + P_\theta(r_0)) \simeq \frac{5 q^2}{18 M^2}$, then
\begin{equation}
\lambda_\star \simeq \frac{1}{3\sqrt{3} M} \left(1 - 4\pi (9 M^2) (\rho + P_\theta) + \frac{3}{2} \left( \frac{3 M q^2}{27 M^3} \right)\right) = \frac{1}{3\sqrt{3} M} \left(1 - \frac{q^2}{9 M^2}\right).
\label{eq:L1Bardeen}
\end{equation}
Additionally, we can compute the shift in the photon sphere radius $\delta r$ by using equation (\ref{eq:deltar}) as
\begin{equation}
\delta r \simeq \frac{3M}{2} \left( -\frac{q^2}{3 M^2} - \frac{6 q^2}{27 M^2} \right) = - \frac{5 q^2}{6 M} \, ,
\end{equation}
which agrees with the WKB analysis of the Bardeen black hole case up to the first order \cite{Konoplya:2023ahd}.  

In particular for this case, $\rho + P_\theta = \frac{15 M q^2 r^2}{8 \pi (q^2 + r^2)^{7/2}}$, then the energy conditions NEC and WEC are satisfied.
Within our perturbative approximation, SEC and DEC reduce to
\begin{align}
\textbf{SEC:}&\quad 
|q| \le \sqrt{\frac{3}{2}}r
\, \overset{\circ}{\Longrightarrow} \, \frac{|q|}{M} \leq \frac{3\sqrt{6}}{5} ,
\label{eq:SEC1} \\
\textbf{DEC:}&\quad 
|q| \ge \frac{r}{2}
\, \overset{\circ}{\Longrightarrow} \, \frac{|q|}{M} \geq \frac{3}{5} (\sqrt{14}-2),
\label{eq:DEC1}
\end{align}
where $\overset{\circ}{\Longrightarrow}$ implies ``around UCOP". While NEC, WEC, and SEC are compatible with $|q|/M \ll 1$, DEC is not compatible with a small $|q|/M$.

\begin{figure}[htbp]
  \centering
  % Top row: two side-by-side figures
  \begin{subfigure}[t]{0.49\textwidth}
    \includegraphics[width=\textwidth]{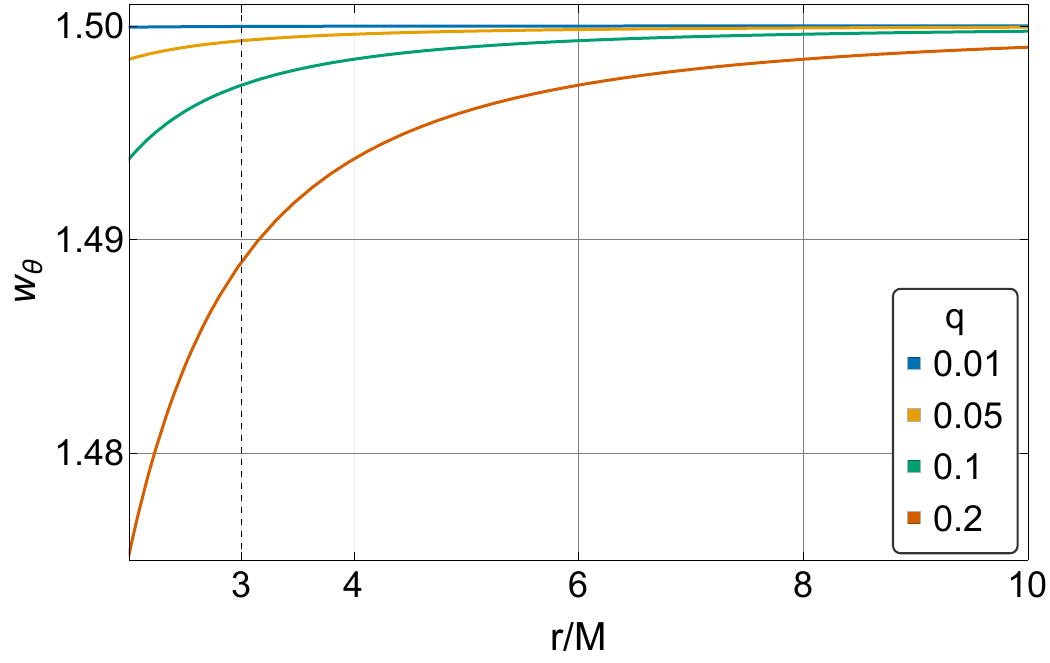}
    \caption{Tangential pressure parameter $w_{\theta}(r)$ as a function of radius for various values of the charge-like parameter $q$.}
    \label{fig:fig1-1}
  \end{subfigure}
  \hfill
  \begin{subfigure}[t]{0.49\textwidth}
    \includegraphics[width=\textwidth]{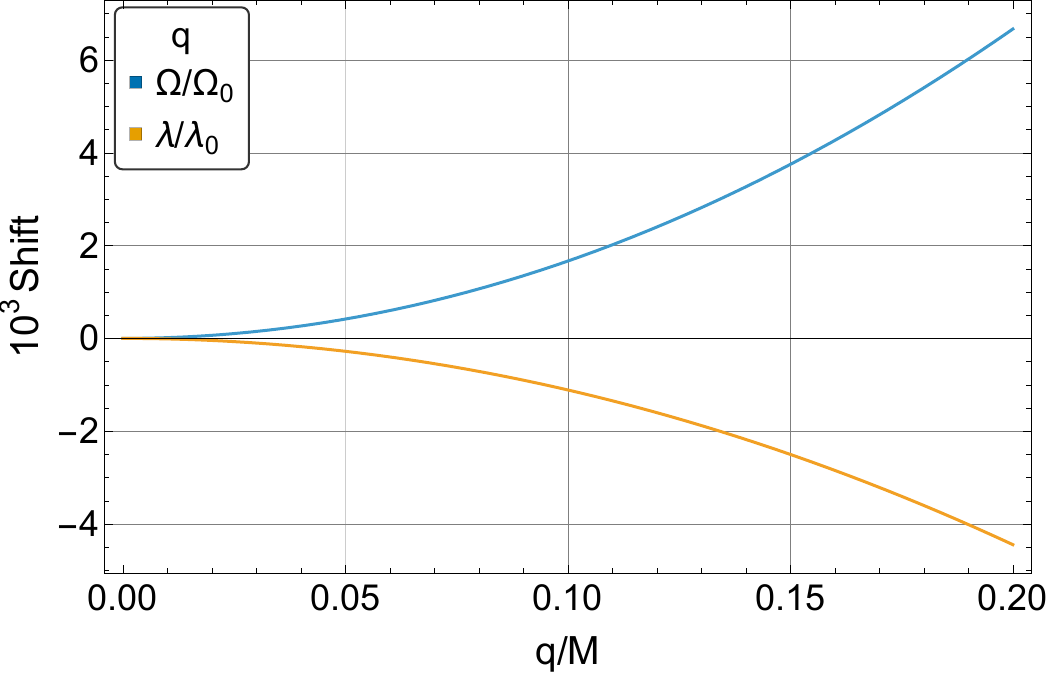}
    \caption{
    Relative shifts of the QNM frequency $\delta \Omega/\Omega_0$ and damping rate $\delta\lambda/\lambda_0$ as functions of $q/M$.} 
    \label{fig:fig1-2}
  \end{subfigure}
  \caption{
    Tangential pressure parameter $w_\theta(r)$ and leading eikonal QNM shifts for the Bardeen black hole.}
  \label{fig:bardeen-static-shifts}
\end{figure}

In Fig.~\ref{fig:fig1-1}, we observe that the tangential fluid parameter $w_\theta$ becomes smaller if we increase the value of the hairy parameter $q$. At the asymptotic region, the value of the tangential parameter tends to $1.5$ for any finite value of $q$.
In Fig.~\ref{fig:fig1-2} we can observe that the $\delta \Omega / \Omega_0$ is positive and $\delta \lambda / \lambda_0$ negative as expressed in \eqref{eq:O1} and \eqref{eq:L1Bardeen}. Where $\delta \Omega / \Omega_0$ would be positive and $\delta \lambda / \lambda_0$ would be negative for any value of $q$. Furthermore, equation~\eqref{eq:rho1} shows that the mass density $\rho$ remains positive for all values of $q$. 

\subsection{Hayward black hole}

The metric function of the Hayward black hole is given by 
\begin{equation}
 f(r)=1/h(r)= 1-\dfrac{2M r^2}{r^3+ q^3} \,,
\end{equation}
where $q$ is a constant. The standard Hayward metric is often written as
\begin{equation}
 f(r)=1-\frac{2Mr^2}{r^3+2M\ell_{\rm H}^2}\,,
\end{equation}
where $\ell_{\rm H}$ is a positive length scale associated with the regular core. Therefore, in our notation, $q^3=2M\ell_{\rm H}^2$.
Since $M>0$ and $\ell_{\rm H}^2\ge0$, the 
% physical 
standard
Hayward branch corresponds to $q \ge 0$.
%\medskip 
Through the Einstein equations, we can write down the energy density and pressures in terms of $M$ and $q$ as 
\begin{equation}
\rho = - P_r =  \dfrac{3 q^3 M}{4 \pi (r^3+q^3)^2} \,, \quad 
P_\theta = - \dfrac{3q^3M(q^3-2r^3)}{4 \pi (r^3+q^3)^3} \,.  
 \label{eq:rho2}
\end{equation}
We can read off the state parameters as
\begin{equation}
 w_r = -1 \,, \quad w_\theta = - \dfrac{q^3 -2r^3}{q^3+r^3} \,. 
\end{equation}

In the scenario where the extra-parameter $q^3$ is sufficiently small, i.e., $|q/M|\ll 1$, around the UCOP $r_\star$, we estimate $\delta f(r)$ $\delta f(r)$ by expanding $f(r)$ up to first order
\begin{equation}
f(r) \simeq 1 - \frac{2M}{r} \left( 1 - \frac{q^3}{r^3} \right)\, \Longrightarrow \, \delta f (r) \simeq \frac{2 M q^3}{r^4} .
\end{equation}
We can calculate the quasinormal mode components. By keeping terms up to $q^3$, we obtain $4 \pi r_0^2 (\rho + P_\theta) \simeq \frac{q^3}{9M^3}$ and
\begin{equation}
\Omega_\star \simeq 
%\frac{1}{3\sqrt{3} M} \left( 1 + \frac{3}{2}\left(\frac{2 M q^3}{81 M^4}\right) \right) \,=\, 
\frac{1}{3\sqrt{3} M} \left(1+\frac{q^3}{27 M^3} \right)\, , \quad
\lambda_\star \simeq 
%\frac{1}{3\sqrt{3} M} \left(1 - 4\pi (9 M^2) (\rho + P_\theta) + \frac{3}{2} \left( \frac{3 M l^2}{27 M^3} \right)\right) = 
\frac{1}{3\sqrt{3} M} \left(1 - \frac{2 q^3}{27 M^3}\right).
\label{eq:LO2}
\end{equation}
Additionally, we can compute the shift in the photon sphere radius $\delta r$ by using equation (\ref{eq:deltar}) as
\begin{equation}
\delta r \simeq -\frac{2 q^3}{9 M^2} \,.
\end{equation} 

\begin{figure}[htbp]
  \centering
  % Top row: two side-by-side figures
  \begin{subfigure}[t]{0.48\textwidth}
     \includegraphics[width=\textwidth]{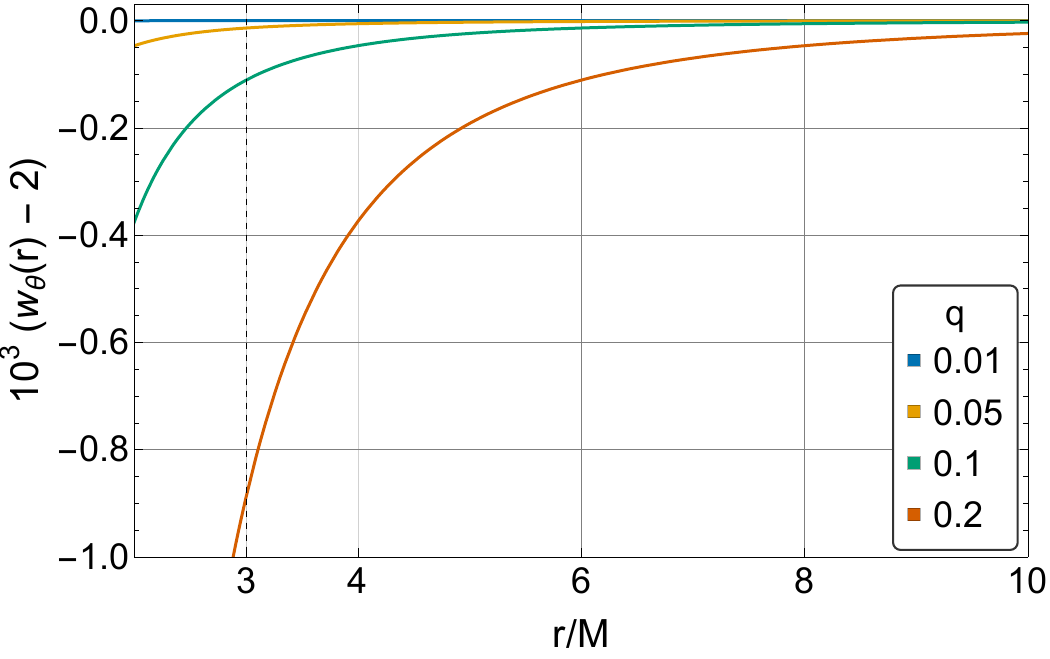}
    \caption{Tangential pressure parameter $w_\theta(r)$ as a function of radius for various values
    of the charge-like parameter $q$.}
    \label{fig:fig2-1}
  \end{subfigure}
  \hfill
  \begin{subfigure}[t]{0.48\textwidth}
    \includegraphics[width=\textwidth]{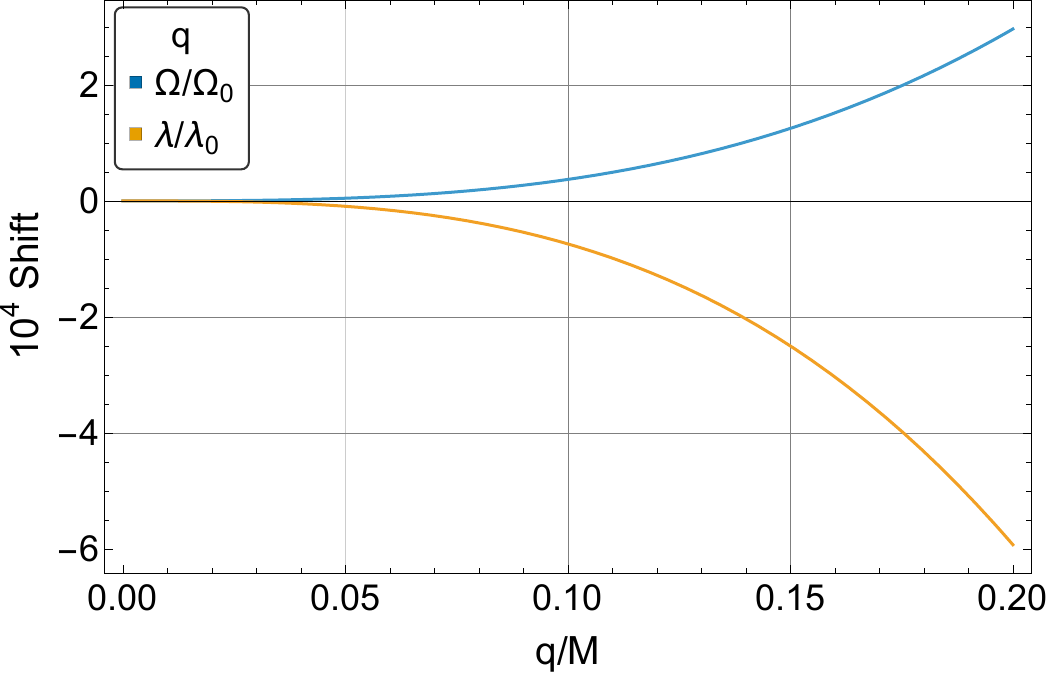}
    \caption{
    % Shifts in quasinormal-mode frequencies due to anisotropic fluid hair.
    Relative shifts of the QNM frequency $\delta \Omega/\Omega_0$ and damping rate $\delta\lambda/\lambda_0$ as functions of $q/M$.
    }
    \label{fig:fig2-2}
  \end{subfigure}

  \caption{
  Tangential pressure parameter $w_\theta(r)$ and leading eikonal QNM shifts for the Hayward black hole.
  % Representative QNM and geometric shifts for the anisotropic fluid halo Hayward model.
  }
  \label{fig:hayward-static-shifts}
  
\end{figure}
In particular for this case, $\rho + P_\theta = \frac{9 M q^3 r^3}{4 \pi (q^3 + r^3)^{3}}$, then the energy conditions NEC and WEC are satisfied when $q > 0$. SEC and DEC reduce to
\begin{align}
\textbf{SEC:}&\quad 
 0 \le q \le \sqrt[3]{2}\, r
\, \overset{\circ}{\Longrightarrow} \,
0 \le \frac{q}{M} \le 
\bigg({\frac{27}{4}+\frac{\sqrt{189}}{2}}\bigg)^{1/3}
+
\bigg({\frac{27}{4}-\frac{\sqrt{189}}{2}}\bigg)^{1/3}\,,
\label{eq:SEC2} \\[4pt]
\textbf{DEC:}&\quad 
q > \frac{r}{\sqrt[3]{2}}
\, \overset{\circ}{\Longrightarrow} \,
\frac{q}{M} >
\bigg({\frac{27+\sqrt{837}}{4}}\bigg)^{1/3}
+
\bigg({\frac{27-\sqrt{837}}{4}}\bigg)^{1/3}.
\label{eq:DEC2}
\end{align}
As in the previous case, NEC, WEC, and SEC are compatible with $|q|/M \ll 1$, and the Dominant condition is not compatible with a small $|q|/M \ll 1$.

In Fig.~\ref{fig:fig2-1}, we observe that the tangential fluid parameter $w_\theta$ becomes smaller at the UCOP if we increase the value of the hairy parameter $q$. At the asymptotic region, the value of the tangential parameter tends to $2$ for any finite value of $q$.
In Fig.~\ref{fig:fig2-2} we can observe that the $\delta \Omega / \Omega_0$ is positive and $\delta \lambda / \lambda_0$ negative as expressed in \eqref{eq:LO2}. 

\subsection{Kiselev black hole}

Quasinormal modes of Kiselev-type black holes surrounded by quintessence have been studied previously, for example, by Chen and Jing~\cite{ChenJing:2005}, who computed scalar-field QNMs using WKB methods and analysed the dependence on the quintessence equation-of-state parameter. The relation between the photon sphere, shadow radius, and QNMs has also been studied for black holes with quintessence-like matter, including Kiselev-type geometries~\cite{YuChenGao:2022}. QNMs of black holes surrounded by anisotropic matter fields have also been studied using explicit perturbation equations for scalar and electromagnetic fields~\cite{SagarJC:2025}. These results provide useful context for the Kiselev example analysed in this subsection.

\medskip

The Kiselev solution \cite{Kiselev:2002dx} describes static spherically symmetric black holes with quintessential matter distribution. 
The metric function is given by 
\begin{equation}
 f(r)=1/h(r)= 1-\dfrac{2M}{r} - \dfrac{k}{r^{1+3\,w_q}} \,,
\end{equation}
where $k$ controls the strength of the surrounding matter field and $w_q$ is an effective state parameter. In the original quintessence interpretation, one usually considers the range
$ -1 < w_q < -1/3$, which is associated with accelerated expansion.

For later comparison with the energy-condition discussion, we also note that the Kiselev stress tensor and its physical interpretation have been analysed in detail in Refs.~\cite{Visser:2019brz,Boonserm:2019phw}.

From the Einstein equations, the associated energy density and pressures are obtained as

\begin{equation}
\rho = - P_r = -\dfrac{3 k \, w_q}{8 \pi \, r^{3(1+w_q)}}\,, \quad 
P_\theta = -\dfrac{3 \, k \, w_q (1 + 3 \, w_q)}{16 \pi \, r^{3(1+w_q)}} \,.  
\end{equation}
%%%
Accordingly, the state parameters are
\begin{equation}
 w_r = -1 \,, 
 \qquad 
 w_\theta = \dfrac{1+3 \, w_q}{2} \,.
\end{equation}
Since the source is anisotropic, we define the averaged pressure
\begin{equation}
 P_q := \frac{P_r+2P_\theta}{3} = w_q \rho  \,,
\end{equation}
which shows that $w_q$ can be interpreted as the effective state parameter of the anisotropic source.

If we treat $|k|$ as a small parameter, we can treat the last term of the metric function $f(r)$ as a small deviation from the Schwarzschild metric function
\begin{equation}
\delta f (r) = -\frac{k}{r^{1+3 w_q}} .
\end{equation}
The frequency components are
\begin{equation}
\Omega_\star \simeq \frac{1}{3\sqrt{3} M} \left( 1 -\frac{3 k}{2 (3M)^{1+3 w_q}} \right)\,, \quad \lambda_\star \simeq \frac{1}{3\sqrt{3} M} \left(1 + \frac{ 3 w_q (1 + w_q) - 2}{4 (3^{3 w_q} M^{1+3 w_q})} k \right).
\end{equation}

\begin{figure}[htbp]
  \centering
  % Top row: two side-by-side figures
    \begin{subfigure}[t]{0.31
    \textwidth}
    \includegraphics[width=\textwidth]{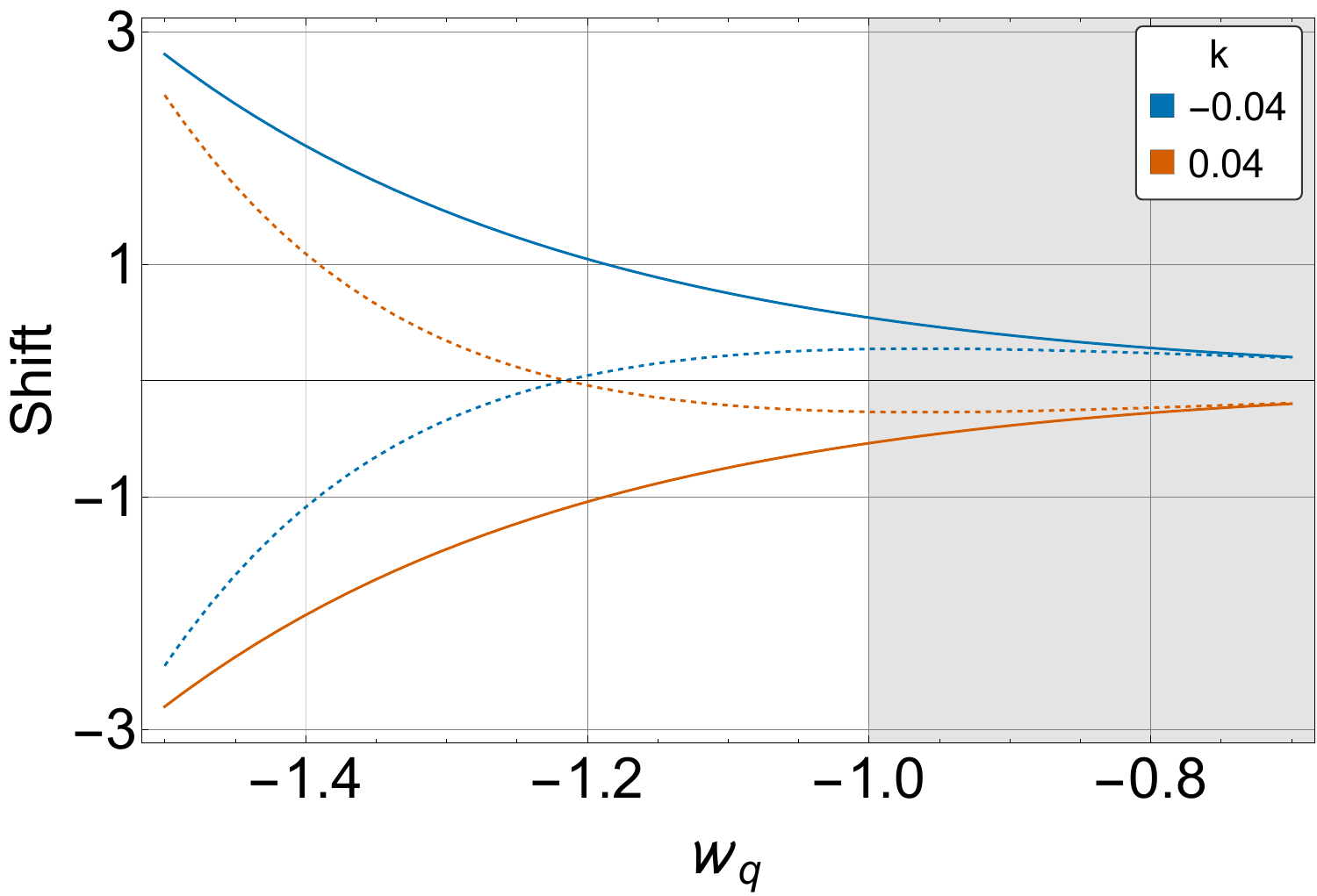}
    \caption{}
    \label{fig:fig3-1}
    \end{subfigure}
  \hfill
  \begin{subfigure}[t]{0.325\textwidth}
    \includegraphics[width=\textwidth]{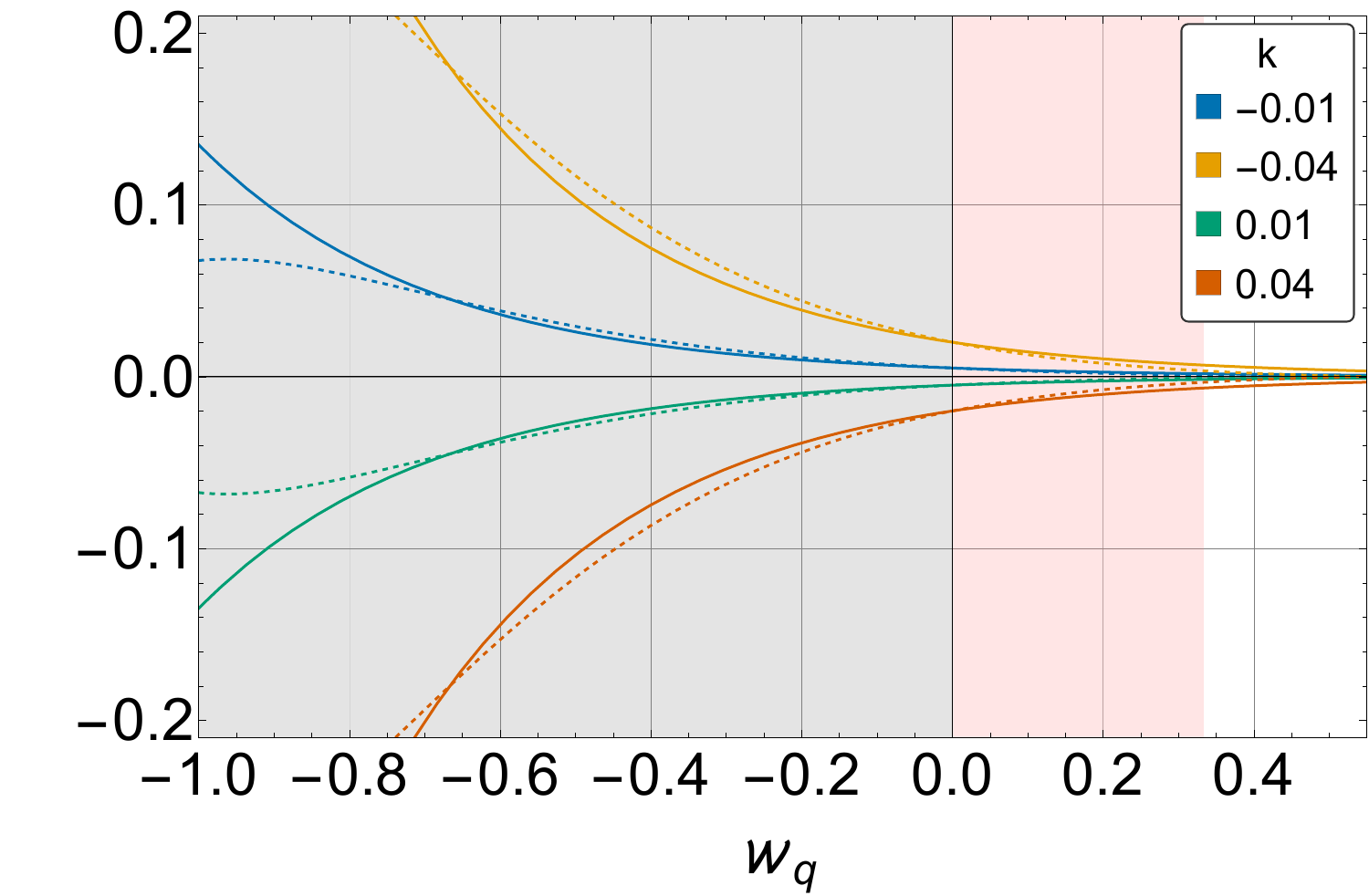}
    \caption{}
    \label{fig:fig3-2}
  \end{subfigure}
  \hfill
  \begin{subfigure}[t]{0.325
  \textwidth}
    \includegraphics[width=\textwidth]{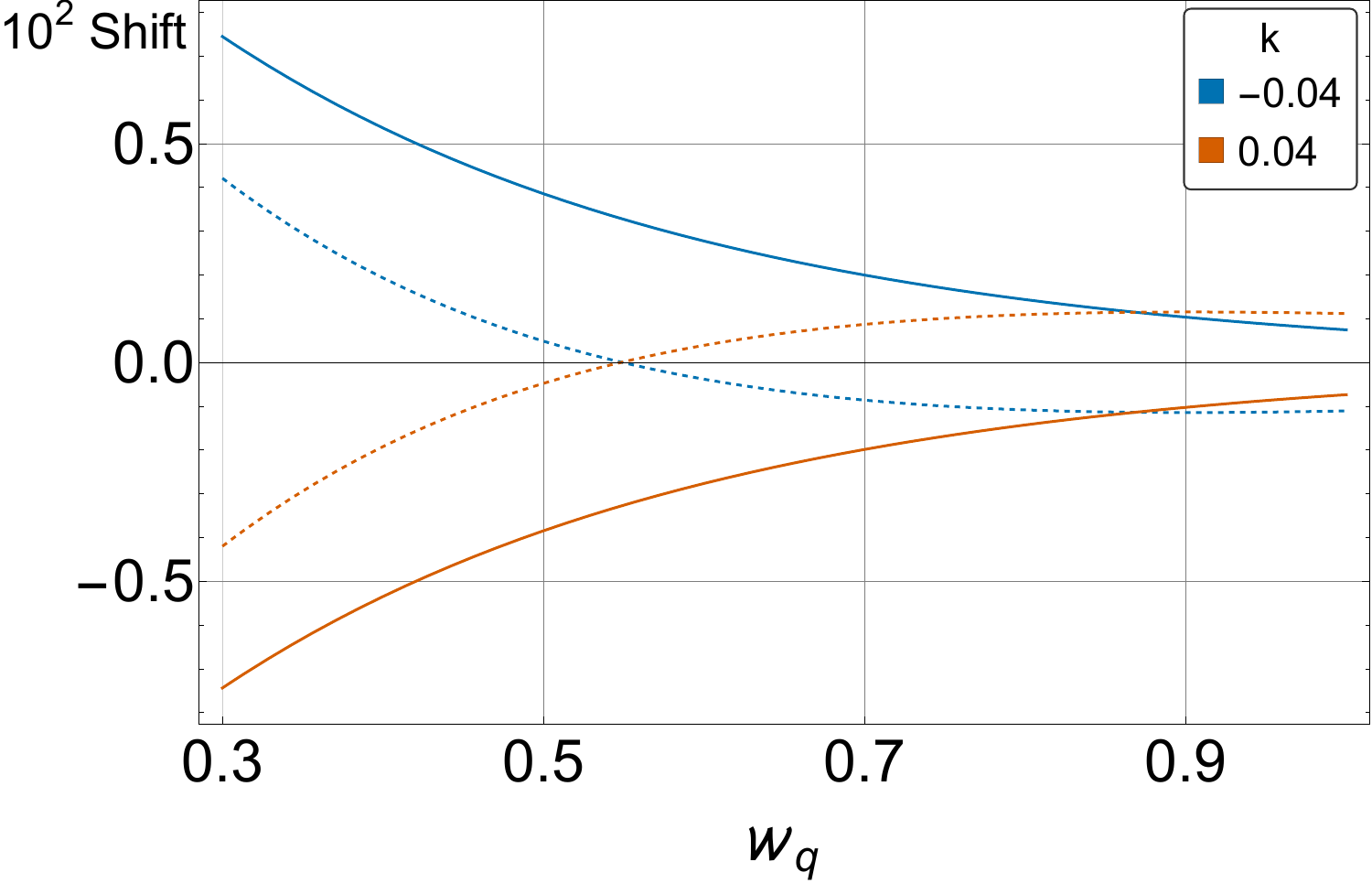}
    \caption{ }

    \label{fig:fig3-3}
  \end{subfigure}
  \caption{QNM shifts for different values of the Kiselev equation-of-state parameter $w_q$. Continuous lines denote $\delta\Omega / \Omega_0$, corresponding to the leading shift of the eikonal oscillation frequency, while dashed lines denote $\delta\lambda / \lambda_0$, corresponding to the leading shift of the damping rate. Different curves correspond to different values of the matter-strength parameter $k$. The shaded regions indicate the parameter intervals satisfying the DEC, the most restrictive energy condition considered in this work: grey for the $k>0$ case and pink for the $k<0$ case.
  }
  \label{fig:Kiselev1}
\end{figure}

Different $w_q$ values tell how fast the ``hair" $\delta f (r)$ decays (or grows). In its original construction, the source is ``quintessence-like", giving a negative tangential pressure. 
For $w_q = -1$, the perturbation term behaves as the cosmological constant, and both pressures are equal.
%For $w_q<-1$, makes $\delta f$ large, despite $k$. The tangential pressure becomes larger in magnitude than the radial pressure.
For values larger than $-1/3$, the hair $\delta f$ decays rapidly as $r$ grows, giving us a metric more similar to the Schwarzschild case. 
In particular, for the value $w_q = 1/3$, if we consider $k = -Q^2$ as charge, the metric reduces to the Reissner-Nordström model. In this last case, tangential and radial pressure are equal in magnitude but opposite in sign. 
For the $k<0$ scenario, in Figs.~\ref{fig:fig3-1} and \ref{fig:fig3-3}, we can observe that in certain ranges of $w_q$, the behaviour of the QNM components resembles the Bardeen (Fig.~\ref{fig:fig1-2}) and Hayward (Fig.~\ref{fig:fig2-2}) scenarios: $\delta\Omega/\Omega_0$ becomes positive and $\delta\lambda/\lambda_0$ turns negative. In contrast, the ``quintessence" regime, $w_q$  from $-1$ to $-1/3$, has a different behaviour as seen in Fig.~\ref{fig:fig3-2}. We observe that the sign depends on $k$, but regardless of the sign of $k$, both shifts have the same sign. 
In this model, $w_\theta$ depends only linearly on $w_q$.

\begin{figure}[htbp]
  \centering
  % Top row: two side-by-side figures
  \begin{subfigure}[t]{0.47\textwidth}
    \includegraphics[width=\textwidth]{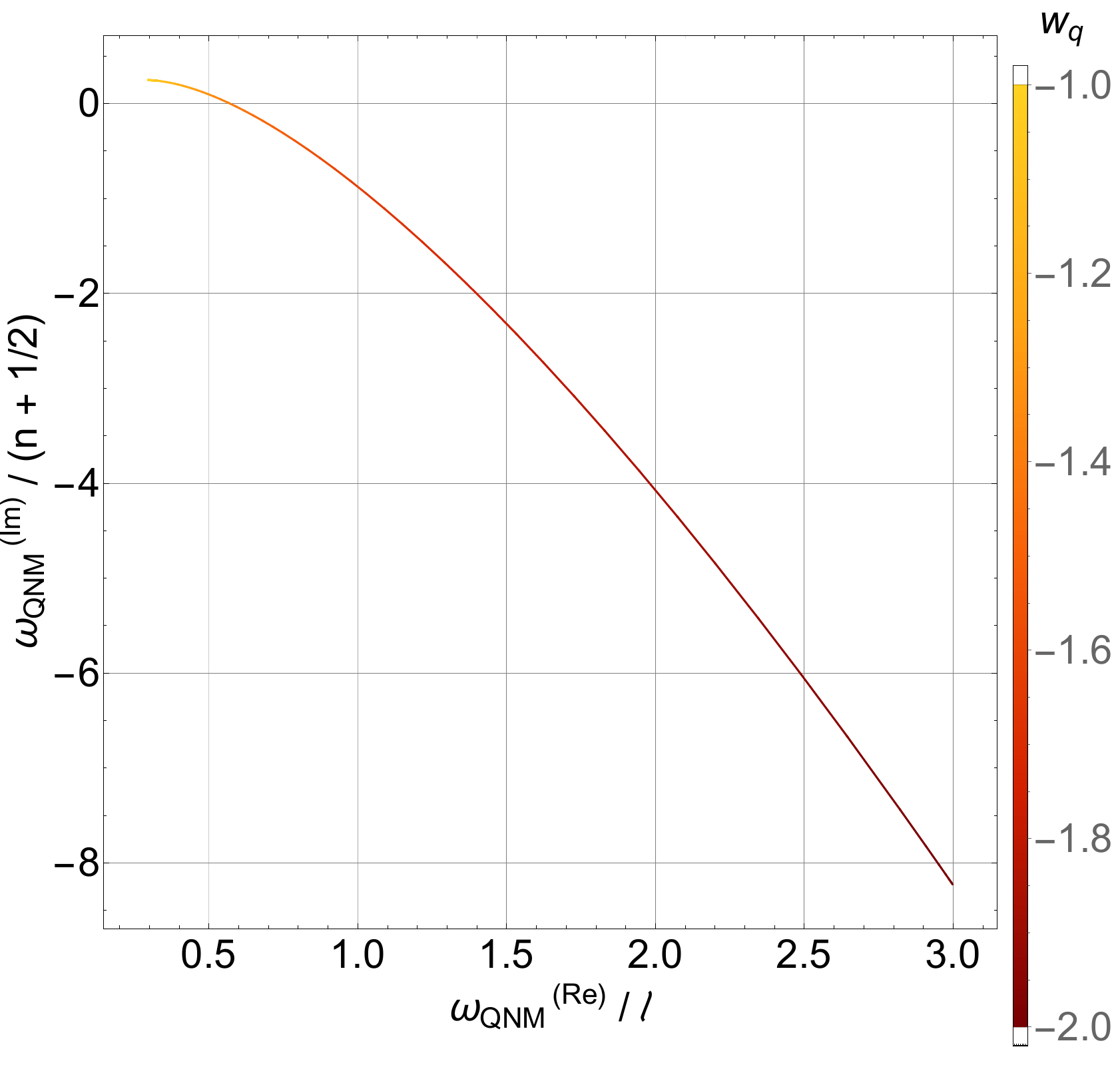}
    \caption{Trajectory in the complex plane of a QNM frequency for $k = -0.04$  evaluated across $[-2,-1]$ values of the equation-of-state parameter \(w_q\).}
    \label{fig:fig3-4}
  \end{subfigure}
  \hfill
  \begin{subfigure}[t]{0.47\textwidth}
    \includegraphics[width=\textwidth]{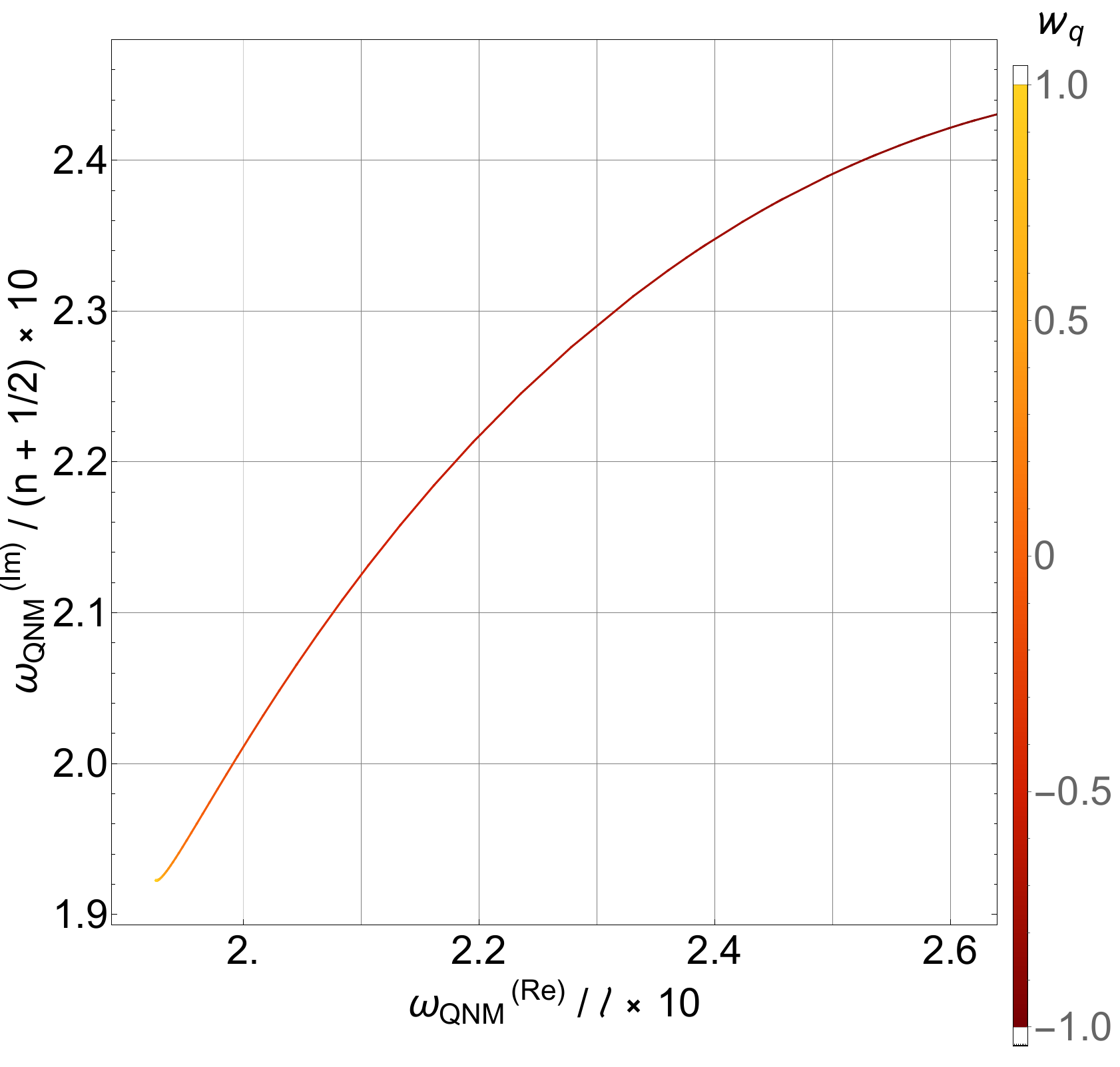}
    \caption{Trajectory in the complex plane of a QNM frequency for $k = -0.04$  evaluated across $[-1,1]$ values of the equation-of-state parameter \(w_q\).}
    \label{fig:fig3-5}
  \end{subfigure}
  \vskip \baselineskip
  \begin{subfigure}[t]{0.47\textwidth}
    \includegraphics[width=\textwidth]{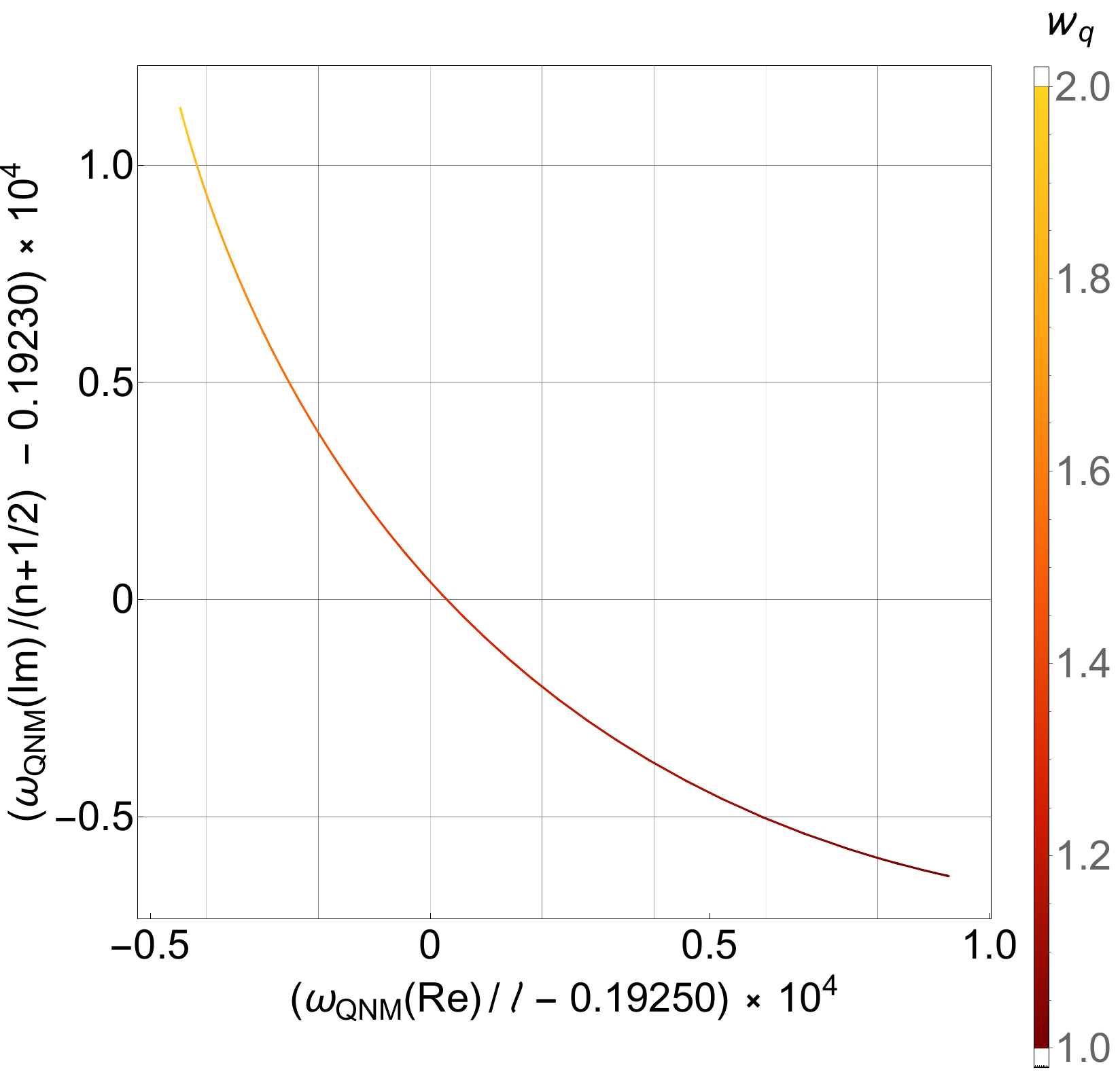}
    \caption{Trajectory in the complex plane of a QNM frequency for $k = -0.04$  evaluated across $[1,2 ]$ values of the equation-of-state parameter $w_q$.}
    \label{fig:fig3-6}
  \end{subfigure}
  \hfill
  \begin{subfigure}[t]{0.48\textwidth}
    \includegraphics[width=\textwidth]{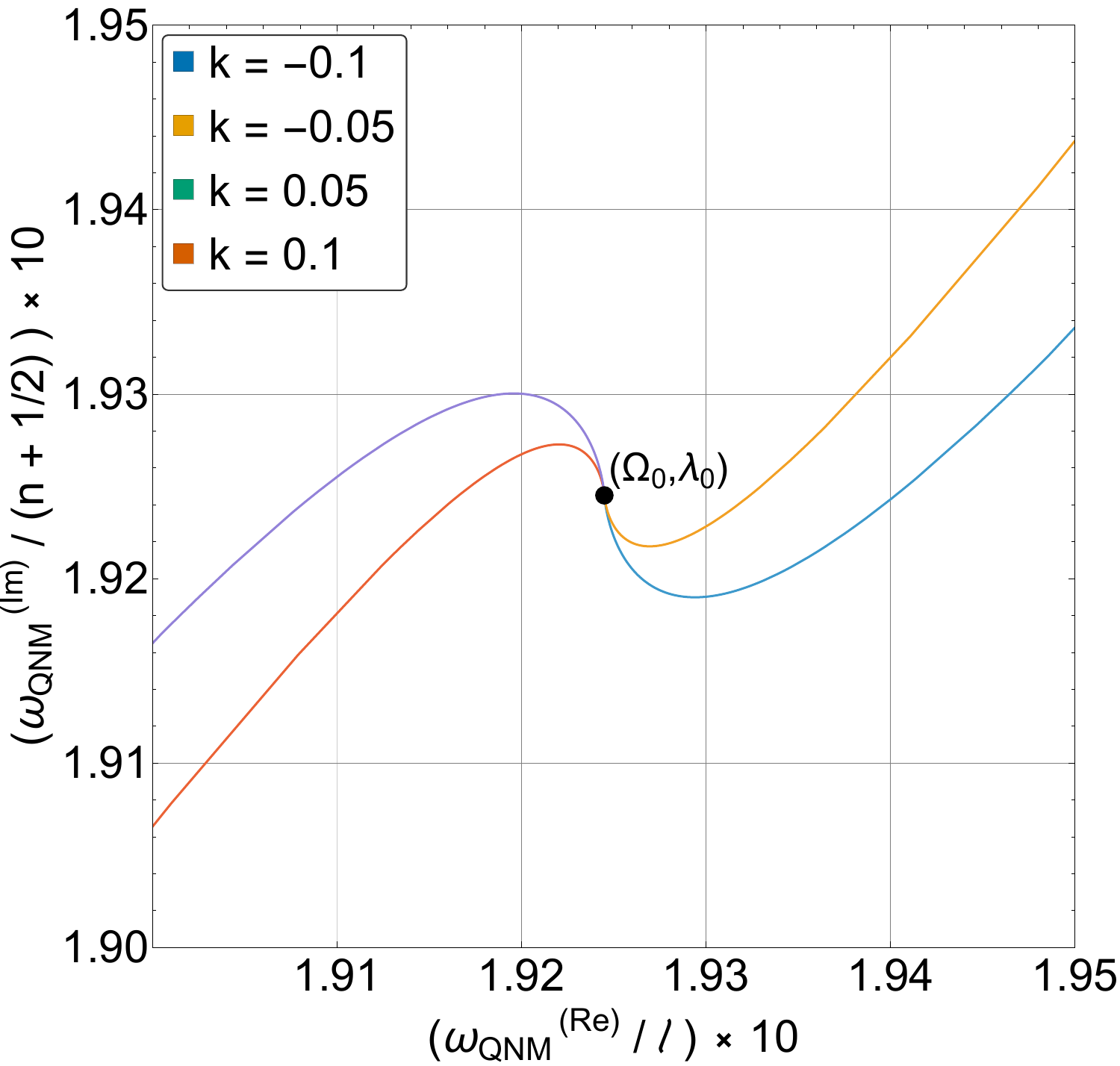}
    \caption{Trajectory of QNM frequencies in the complex plane under varying anisotropic fluid parameter $k={-0.10,\,-0.05,\,0.05,\,0.10}$, illustrating how the real and imaginary components of the QNM frequency shift relative to the reference point $\{\Omega_0,\lambda_0\}$ at $k = 0$.}

    \label{fig:fig3-7}
  \end{subfigure}
  \caption{Representative QNM and geometric shifts for the anisotropic fluid halo Kiselev model. The horizontal axis represents the oscillation-frequency component, $\mathrm{Re}(\omega_{\rm QNM})/\ell$, while the vertical axis represents the damping-rate component, $-\mathrm{Im}(\omega_{\rm QNM})/(n+1/2)$. The trajectories show how varying the equation-of-state parameter $w_q$ or the matter-strength parameter $k$ moves the mode in the complex-frequency plane.}
  \label{fig:kiselev-qnm-trajectories}
\end{figure}

\begin{figure}[htbp]
  \centering
  \begin{subfigure}[t]{\textwidth}
    \includegraphics[width=\textwidth]{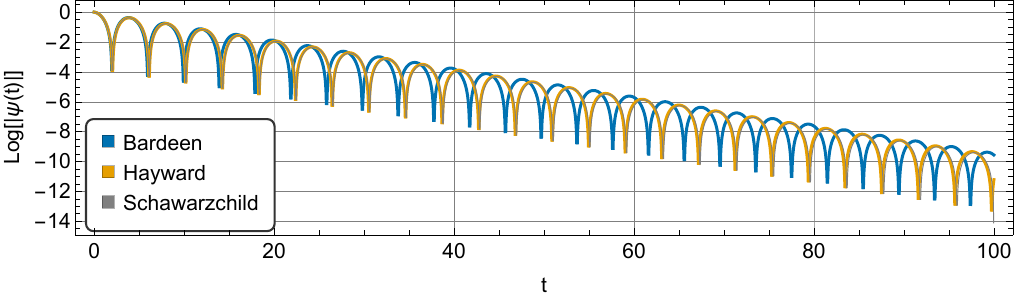}
    \caption{
    % Quasinormal absolute amplitude 
    Damping oscillation waveforms for the Bardeen, Hayward and Schwarzschild cases with $q=0.4$.}
    \label{fig:figall1}
  \end{subfigure}

    \vskip\baselineskip
  \begin{subfigure}[t]{\textwidth}
    \includegraphics[width=\textwidth]{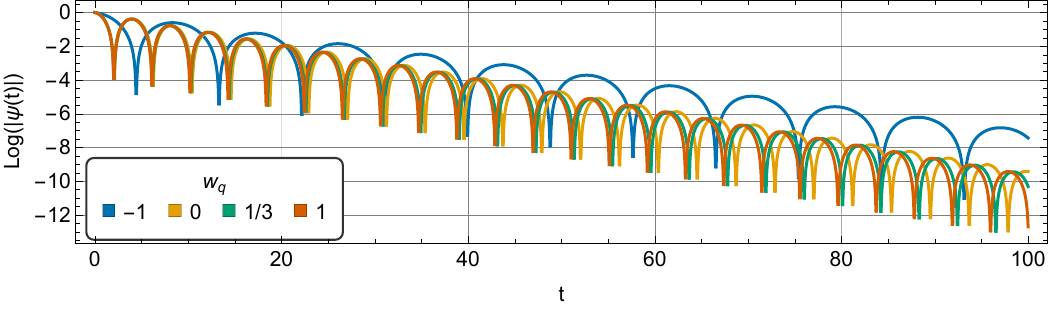}
    \caption{Damping oscillation waveforms for the Kiselev model with $k=0.04$ for different values of $w_q$.}
    \label{fig:figall2}
  \end{subfigure}
  \caption{
  Time evolution of the illustrative ringdown waveform $\Psi(t)=\Re\!\left[e^{-i\,\omega_{\rm QNM} t}\right]$. We compare the amplitudes within the different models for the quasinormal mode $n=0$ and $\ell=4$. Although $\ell=4$ may not appear to be a large enough value to justify the eikonal approximation, the existing literature~\cite{Iyer:1986nq,Berti:2005eb} has shown that the eikonal/WKB description can already give good quantitative agreement for comparable black-hole QNM calculations. These plots are intended as illustrative waveforms constructed from the geodesic/eikonal estimates, not as full low-$\ell$ gravitational waveforms.
}
  \label{fig:three-panel-mixed-3}
\end{figure}

Fig.~\ref{fig:kiselev-qnm-trajectories} shows trajectories describing particular cases of the QNMs in the complex plane normalized by their modes. In Figs.~\ref{fig:fig3-4}, \ref{fig:fig3-5} and \ref{fig:fig3-6}, the color changes mark the different values of $w_q$. We can observe completely different behaviours. In panel~\ref{fig:fig3-7}, we can observe a comparison for different values of $k$. The quantities shown in Figs.~\ref{fig:kiselev-qnm-trajectories} and ~\ref{fig:three-panel-mixed-3} are constructed from the geodesic quantities $\Omega$ and $\lambda$ derived above, using the established eikonal QNM--geodesic correspondence.

Additionally, we can compute the shift in the photon sphere radius $\delta r$ by using equation (\ref{eq:deltar}) as
\begin{equation}
\delta r \simeq \frac{3 k (1+ w_q)}{2 (3M)^{3 w_q}} \,.
\end{equation}

In particular for this case, $\rho + P_\theta = -\frac{9 \, k \, w_q (1+w_q)}{16 \pi r^{3(1+w_q)}}$, then the energy conditions reduce to
\begin{align}
\textbf{NEC:}\quad
&k\,w_q(1+w_q)\le 0
\;\;\overset{\circ}{\Longrightarrow}\;\;
\begin{cases}
k>0: & w_q\in\left[-1,\,0\right],\\[2pt]
k<0: & w_q\notin\left(-1,\,0\right),
\end{cases}
\label{eq:NEC3}\\[6pt]
\textbf{WEC:}\quad
&k\,w_q\le 0 \;\;\land\;\; 1+w_q\ge 0
\;\;\overset{\circ}{\Longrightarrow}\;\;
\begin{cases}
k>0: & w_q\in\left[-1,\,0\right],\\[2pt]
k<0: & w_q\ge 0,
\end{cases}
\label{eq:wEC3}\\[6pt]
\textbf{SEC:}\quad
&k\,w_q(1+w_q)\le 0 \;\;\land\;\; k\,w_q(1+3w_q)\le 0
\;\;\overset{\circ}{\Longrightarrow}\;\;
\begin{cases}
k>0: & w_q\in\left[-\tfrac13,\,0\right],\\[2pt]
k<0: & w_q\notin\left(-1,\,0\right),
\end{cases}
\label{eq:SEC3}\\[6pt]
\textbf{DEC:}\quad
&\bigl|k\,w_q(1+3w_q)\bigr|\le -2k\,w_q
\;\;\overset{\circ}{\Longrightarrow}\;\;
\begin{cases}
k>0: & w_q\in\left[-1,\,0\right],\\[2pt]
k<0: & w_q\in\left[0,\,\tfrac13\right].
\end{cases}
\label{eq:DEC3}
\end{align}
Unlike the previous cases, the energy conditions are not dependent on the position of the UCOP radii. All conditions are compatible with a small $|k|$.

In Figs.~\ref{fig:figall1} and \ref{fig:figall2}, we demonstrate the damping oscillation waveforms $\log{(|\Psi|)}$ for Schwarzschild, Bardeen, Hayward, and Kiselev models. 
One can clearly see the decay of the amplitude in time, exhibiting the damped oscillation. 
The slope connecting the maximal points in the waveforms describes the damping rate, while the spacing between the cusps represents the oscillation frequency.
Across all models, the QNM amplitude exhibits the expected damped ringdown, but the details of the decay differ noticeably between cases. In Fig.~\ref{fig:figall1}, we can observe how, in comparison to the vacuum scenario, signals for Bardeen and Hayward travel with a similar decaying rate, but with a clearly different frequency. 
In Fig.~\ref{fig:figall2}, we can observe different signals for the Kiselev model according to the $w_q$ value. There is a clear difference in the 
% envelope amplitudes. 
slope connecting the maximal points. 
In particular, the case $w_q = -1$ has 
% a wider amplitude, 
a gentle downward slope, and therefore a smaller decay rate. 
On the other hand, the cusps are more widely spaced in comparison with the other cases. It represents a slower oscillation with a slower decay. 
% as it travels in the rad

\section{Stationary rotating hairy black hole and QNMs} \label{ch:rotating}

In this section, we consider stationary rotating hairy black holes. As discussed in the Introduction, the QNM--photon-orbit correspondence in rotating spacetimes is more subtle than in the static, spherically symmetric case, since rotation breaks the degeneracy in the azimuthal harmonic number $\mathbf{m}$. Here, we restrict the analysis to null rays trapped in the equatorial plane. This corresponds to the eikonal sector with $\ell=|\mathbf{m}|$, describing the co-rotating and counter-rotating equatorial branches~\cite{Berti:2005eb,Berti:2005ys,Dolan:2010wr,Yang:2012he}. The co-rotating and counter-rotating branches are labelled by $\varepsilon=-1$ and $\varepsilon=+1$, respectively, and our sign convention gives $\mathbf{m}=-\varepsilon\ell$. In this restricted sector, the QNM frequencies are estimated as
\begin{equation}
\omega_{\rm QNM}^{(\varepsilon)}
= \mathbf{m}\Omega_{\varepsilon}
-i\left(n+\frac{1}{2}\right)\lambda_{\varepsilon}.
\end{equation}

Our sign conventions are summarised in Table~\ref{tab:signs}.

\begin{table}[ht]
\centering
\caption{Sign and range conventions for the rotating-case analysis.}
\begin{tabular}{l|lllllll}
\toprule
\textbf{} & \textbf{$\varepsilon$} & \textbf{$a$} & \textbf{$\Omega_\varepsilon$} & \textbf{$\lambda_\varepsilon$} & \textbf{$\delta{\Omega_{\varepsilon}}$} & \textbf{$\delta{\hat{\Omega}_{\varepsilon}}$} & \textbf{$\delta{\lambda_{\varepsilon}}$} \\
\midrule
\textbf{Co-rotating} & $-$ & $+$ & $+$ & $+$ & $+$& $+$ & $\pm$ \\
\textbf{Counter-rotating} & $+$ & $+$ & $-$ & $+$ & $-$& $+$ & $\pm$ \\
\bottomrule
\end{tabular}
\label{tab:signs}
\end{table}

\subsection{Rotating Hairy Black Hole and null geodesics} \label{subsec:rotating:formulas}

We focus on the following form of the metric
\bena
ds^2&=& -f dt^2-2a\sin^2\theta (1- f)dtd\varphi + \left\{\Sigma +(2-f)a^2\sin^2\theta \right\} \sin^2\theta d\varphi^2 
\non \\
 &{}& \qquad 
+ \dfrac{\Sigma}{\Sigma f +a^2 \sin^2\theta} dr^2 + \Sigma d\theta^2 \,,
\non
\\
 &{}& f :=1-\dfrac{2m(r)r}{\Sigma} \,, \quad \Sigma :=r^2+a^2 \cos^2\theta \,, 
\label{rotating:metric}
\eena
where \(m(r)\) is an arbitrary differentiable function of \(r\), and \(a\) denotes the angular momentum parameter. For $m(r) = M$, this metric reduces to the Kerr solution.  
The above metric can be obtained from the static metric~(\ref{ansatz:stat:metric}) by setting $f(r)=1/h(r)$ and performing the Newman--Janis transformation~\cite{Newman:1965tw,Drake:1998gf}.
This class of metrics encompasses a fairly large class of stationary rotating hairy black holes, including the rotating Bardeen, the rotating Hayward, the rotating Kiselev, the quantum improved Kerr~\cite{quantumimprovedregularkerr}, among others.

For the rotating metrics generated through the Newman--Janis procedure, the matter interpretation should be treated with some care. Once the metric is specified, an effective stress-energy tensor can be reconstructed from $G_{\mu\nu}=8\pi T_{\mu\nu}^{\rm eff}$. However, this tensor need not correspond to the same matter model as the static seed solution. Thus, in the rotating case, $\rho$, $P_r$, and $P_\theta$ should be understood as effective source variables of the resulting geometry. The energy-condition analysis is therefore a diagnostic of the effective geometry-source pair, not a proof of a unique underlying matter model.

In this class of metrics, the rotating counterpart of a given static solution is obtained by replacing the mass function $m(r)$ by the corresponding static mass profile defined in Eq.~\eqref{eq:metric-split}.

Let us consider circular geodesic orbits. We restrict our attention to the equatorial plane $\theta = \pi/2$. 
The Lagrangian is given as 
\bena
 {\cal L} = \dfrac{1}{2}\left( -f  \dot{t}^2 + \dfrac{r^2 \dot{r}^2}{a^2+r^2f} - 2a(1-f)\dot{t}\dot{\varphi} + [r^2+ (2-f)a^2]\dot{\varphi}^2 \right)=\dfrac{\epsilon}{2} \,,
\eena
where $\epsilon = -1$ for the timelike case or $\epsilon=0$ for the null case. 
As for the static case, we define the two conserved quantities $E$ and $L$ as 
\bena
 E:=
 % -p_t 
 - \dfrac{\partial {\cal L}}{\partial \dot{t}} = f \dot{t} +a(1-f)\dot{\varphi} \,, \quad 
 L:= 
 % p_\varphi = 
 \dfrac{\partial {\cal L}}{\partial \dot{\varphi}} = -a (1-f)\dot{t}+[r^2+(2-f)a^2]\dot{\varphi} \,. 
\eena
Then, by solving for $\dot{t}$ and $\dot{\varphi}$, we obtain
\bena
\dot{t} = \dfrac{1}{\triangle} \left\{ \left( r^2+a^2+\dfrac{2m(r)a^2}{r} \right) E-\dfrac{2m(r)a}{r}L \right\} \,, 
\quad 
\dot{\varphi} = \dfrac{1}{\triangle} \left\{ \dfrac{2m(r)a}{r}E + \left( 1-\dfrac{2m(r)}{r} \right) L \right\} \,, 
\eena
where $\triangle:= a^2+r^2f = r^2+a^2-2m(r)r$.
% With $p_r=\partial {\cal L}/\partial \dot{r}= r^2\dot{r}/\triangle$, the Hamiltonian ${\cal H}$ is given as
% \bena
%   {\cal H} = p_t \dot{t}+p_r \dot{r}+p_\varphi \dot{\varphi} - {\cal L} = -E\dot{t}+p_r \dot{r}+ L \dot{\varphi} - \dfrac{\epsilon}{2} \,,
% \eena
% and it yields the radial equation
Then, from the expression of the Lagrangian, we find 
\bena
 \dfrac{1}{2}\dot{r}^2+V(r)=0 \,, \quad 
  V(r):=\dfrac{1}{2r^2}\left[L^2-(r^2+a^2)E^2 - \dfrac{2m(r)}{r}(aE-L)^2-\triangle \epsilon \right] \,. 
\eena

For clarity, let us define
\[
\mathcal{R}(r):=
L^2-(r^2+a^2)E^2
-\frac{2m(r)}{r}(aE-L)^2-\Delta\epsilon \,
\Rightarrow \, V(r)=\frac{\mathcal{R}(r)}{2r^2}, 
\]
so that
\[
V'(r)
=
\frac{\mathcal{R}'(r)}{2r^2}
-\frac{\mathcal{R}(r)}{r^3}.
\]
The conditions for a circular orbit at radius $r_\star$ are $V(r_\star)=0$ and $V'(r_\star)=0$. Therefore, the first condition is equivalent to $\mathcal{R}(r_\star)=0$. Consequently, the second term in $V'(r_\star)$, which comes from differentiating the denominator, vanishes, and the second circular-orbit condition reduces to
$\mathcal{R}'(r_\star)=0$. From these conditions, we have 
\bena
(a^2+r^2)E^2-L^2 +\dfrac{2m}{r}(aE-L)^2 + \triangle \epsilon =0 , \quad 
2E^2 r+ \left( \dfrac{2m'}{r}-\dfrac{2m}{r^2} \right)(aE-L)^2 + \triangle ' \epsilon =0 \,.
\eena
For the null geodesic
% unstable circular 
case ($\epsilon=0$), it is convenient to introduce the impact parameter $D:=L/E$. Then, the two circular-orbit conditions become
\begin{equation}
 a^2-D^2+r^2+\dfrac{2m}{r}(a-D)^2 \circeq 0 \,, \quad 
 2r^2+\left( 2m'-\dfrac{2m}{r}\right)(a-D)^2 \circeq 0 \,,  
\end{equation}
% Then, 
and we find that the impact parameter and the UCOP radii should satisfy
\begin{equation}
 D_{\varepsilon} \circeq a - \varepsilon\sqrt{\dfrac{r^3}{m-rm'}} \,,
 \label{eq:D1}
\end{equation}
\begin{equation}
  r \circeq  (3m(r) - rm'(r)) \dfrac{D_{\varepsilon} - a}{D_{\varepsilon} +a}. 
 \label{eq:r1}
\end{equation}
The second derivative of the potential is given by
\begin{equation}
 V'' \circeq - \dfrac{E^2(D_{\varepsilon} - a)^2}{r^5}\left[ 3(m-rm') + r^2m'' \right] \,.
\end{equation}
From the standard definitions of the orbital frequency and the Lyapunov exponent, we derive
%hereeee
\begin{equation}
 \Omega_{\varepsilon}  \circeq \dfrac{1}{D_{\varepsilon}} \,, \quad 
 \lambda_{\varepsilon}^2 \circeq \left( \dfrac{D_{\varepsilon}+a}{D_{\varepsilon}}\right)^2 \dfrac{\triangle^2}{r^3} \dfrac{3(m-rm')+r^2m''}{[ (D_{\varepsilon}-2a)m-D_{\varepsilon} \\\ r\\\ m']^2}. 
\end{equation}
Note that for the non-rotating vacuum case, i.e., $a=0$, $m= M$, the above formula reduces to the Schwarzschild result $r_{\star,\varepsilon}=r_0=3M$.  

\medskip 

Now by using the Einstein equations, we find (in Appendix~\ref{app2}) that 
\begin{equation}
 m'= -\dfrac{r^2}{2}G^r{}_r = -4\pi r^2 P_r \,, \quad 
 m'' = -r G^\theta{}_\theta = -8\pi r P_\theta \,. 
\end{equation}
Therefore, at the UCOP radius position $r_{\star, \varepsilon}$, we find 
\begin{align}
 \Omega_{\varepsilon} 
 &= D_{\varepsilon}^{-1} =\left( a - \varepsilon\sqrt{\dfrac{{r_{\star, \varepsilon}}^3}{m+4\pi {r_{\star, \varepsilon}}^3 P_r}} \right)^{-1} \,, 
 \label{eq:OmegaR0}
 \quad \\
 \lambda_{\varepsilon} 
 &= \bigg| \dfrac{D_{\varepsilon}+a}{D_{\varepsilon}} \dfrac{\triangle}{{r_{\star, \varepsilon}}^{3/2}}  \dfrac{(3m -4 \pi {r_{\star, \varepsilon}}^3 (2 P_\theta -3 P_r))^{1/2}}{(D_{\varepsilon} - 2a)m+4\pi D_{\varepsilon} \, {r_{\star, \varepsilon}}^3 P_r} \bigg|\,,
 \label{eq:lamdaR0}
\end{align}
where all quantities are evaluated at $r=r_{\star, \varepsilon}$. 

In order to model these general expressions as a small hair perturbation introduced to a vacuum Kerr black hole, we expand the expressions up to the first order
\bena
r_{\star, \varepsilon} \simeq r_{K, \varepsilon} + \delta r_{\varepsilon} \, ,\quad \\
m(r) \simeq M + \delta m(r)\, ,\quad \\
D_{\varepsilon} \simeq D_{K, \varepsilon} + \delta D_{\varepsilon}\, ,\quad \\
\Omega_{\varepsilon} \simeq \Omega_{K, \varepsilon} + \delta \Omega_{\varepsilon}\, ,\quad \\
\lambda_{\varepsilon} \simeq \lambda_{K, \varepsilon} + \delta \lambda_{\varepsilon}\, ,\quad 
\eena
where {$r_{K, \varepsilon}$, $D_{K, \varepsilon}$, $\Omega_{K, \varepsilon}$, $\lambda_{K, \varepsilon}$} are the values of the parameters in the Kerr case for the branch $\varepsilon$.
For the Kerr black hole, the corresponding quantities are given by
\bena
r_{K, \varepsilon} = 2M \left( 1 + \cos{\left(\frac{2}{3}\arccos{\left(-\frac{\varepsilon\, a}{M}\right)} \right)} \right) \, ,\quad \\
D_{K, \varepsilon} = a - \varepsilon\sqrt{\frac{r_{K, \varepsilon}^3}{M}} \, ,\quad \\
\Omega_{K, \varepsilon}=\frac{1}{D_{K, \varepsilon}} \, ,\quad \\
\lambda_{K, \varepsilon} =  \bigg|\frac{\sqrt{3} \, \triangle(D_{K, \varepsilon} + a)}{\sqrt{M} r_{K, \varepsilon}^{3/2} \, D_{K, \varepsilon} \, (D_{K, \varepsilon}-2a)} \bigg| = 
%- \varepsilon 
\frac{\sqrt{3}(r_{K, \varepsilon}-M)}{r_{K, \varepsilon}(r_{K, \varepsilon}+3M)},
\eena
where the allowed radii are constrained by the extremal Kerr limit $|a|=M$. The co-rotating branch ($\varepsilon = -1$) can take values in the interval $[M,3M]$, while the counter-rotating branch lies in the interval $[3M,4M]$.

For the orbital frequency \eqref{eq:OmegaR0}, we obtain
\begin{align}
\delta D_{\varepsilon}
&= \varepsilon \frac{{r_{K, \varepsilon}}^{3/2}}{2\,M^{1/2}}\Big( \frac{\delta m - {r_{K, \varepsilon}} \delta m'}{M}- \frac{3 \delta r_{\varepsilon}}{r_{K, \varepsilon}}  \Big),\label{eq:deltaD_rot}\\
\delta \Omega_{\varepsilon}
&= -\,\Omega_{K, \varepsilon}^2\,\delta D_{\varepsilon}.\label{eq:deltaOmega_rot}
\end{align}
To calculate the shift of the position of the photon orbits radii, we proceed to eliminate $D$ from the null circular-orbit conditions \eqref{eq:D1} and \eqref{eq:r1}. Then we obtain the following
% to obtain an 
implicit UCOP radius equation 
\begin{equation}
F_{\varepsilon}(r; m, m', a) := r^2 \big(3 m - r - r m'\big) +  2\, a\, \varepsilon \, r^{3/2}\sqrt{m - r m'} \circeq 0,
\end{equation}
which reduces to $r(3M - r) + 2 a \, \varepsilon  \sqrt{M r} \circeq 0$ for the Kerr case. Linearizing $F_{\varepsilon}(r_{K, \varepsilon} + \delta r_{\varepsilon}, M+\delta m,\delta m';a) = 0$ gives
\begin{equation}
\delta r_{\varepsilon} \circeq - \frac{ (\partial F_{\varepsilon} /\partial m)_K\,\delta m + (\partial F_{\varepsilon}/\partial m')_K\,\delta m'}{(\partial F_{\varepsilon}/\partial r)_K}\Bigg|_{r=r_{\mathrm{K, \varepsilon}}},
\end{equation}
where the subscript \(K\) indicates that the corresponding quantities are evaluated in the Kerr spacetime, i.e., at \(r=r_{K,\varepsilon}\) with \(m(r)=M\), 
\begin{equation}
\left(\frac{\partial F_{\varepsilon}}{\partial r}\right)_K = \frac{3}{2}\, r_{K, \varepsilon} (M - r_{K, \varepsilon}),\quad
\left(\frac{\partial F_{\varepsilon}}{\partial m}\right)_K = \frac{{r_{K, \varepsilon}} ^2 (r_{K, \varepsilon} + 3\, M)}{2M},\quad
\left(\frac{\partial F_{\varepsilon}}{\partial m'}\right)_K = -\,\frac{{r_{K, \varepsilon}}^3 (r_{K, \varepsilon} - M)}{2M}.
\end{equation}
Then the perturbation on the photon orbit positions is given by 
\begin{equation}
\delta r_{\varepsilon} \simeq -\frac{1}{3M}\left( {r_{K, \varepsilon}}^2\, \delta m' - \frac{{r_{K, \varepsilon}}({r_{K, \varepsilon}} + 3 \, M)}{{r_{K, \varepsilon}} - M}\, \delta m \right)\Bigg|_{r=r_{K, \varepsilon}}. 
\label{eq:deltarK}
\end{equation}
% 
% If we 
Substituting the expression for $\delta r_{\varepsilon}$ \eqref{eq:deltarK} into the expression for $\delta D_{\varepsilon}$~\eqref{eq:deltaD_rot},  we obtain
\begin{equation}
\delta D_{\varepsilon} \simeq - \frac{2 \, \varepsilon}{r_{K, \varepsilon} - M} \sqrt{\frac{{r_{K, \varepsilon}}^3}{M}} \, \delta m \Bigg|_{r=r_{K, \varepsilon}},
\end{equation}
which gives the perturbation of the impact parameter for the UCOPs. The corresponding orbital frequency is then expressed as
\begin{equation}
    \delta \Omega_{\varepsilon} \simeq  \varepsilon \, D_{K, \varepsilon}^{-2} \left(\frac{2}{r_{K, \varepsilon}-M} \sqrt{\frac{{r_{K, \varepsilon}}^3}{M}}\right)\delta m \Bigg|_{r=r_{K, \varepsilon}} \quad.
    \label{eq:deltaOR}
\end{equation}
% Then, b
By expanding up to the first order, we obtain the Lyapunov exponent as follows
% we obtain
\begin{multline}
    \delta(\lambda_{\varepsilon}^2) \simeq \frac{1}{M \, r_{K, \varepsilon}^2 (3M+r_{K, \varepsilon})^3}\left(-2(r_{K, \varepsilon}^3 + 13 M \, r_{K, \varepsilon}^2-21 M^2\,r_{K, \varepsilon} - 9M^3)\delta m \right. \\
    \left. +  2\, r_{K, \varepsilon} (r_{K, \varepsilon}-M)(r_{K, \varepsilon}+3M)(r_{K, \varepsilon}-5M)\delta m' + r_{K, \varepsilon}^2 (r_{K, \varepsilon}-M)^2 (r_{K, \varepsilon} + 3M)\delta m'' \right)\Bigg|_{r=r_{K, \varepsilon}}.
\end{multline}
Finally we obtain the perturbation by using $\delta \lambda = (2 \lambda_{K, \varepsilon})^{-1}\delta (\lambda^2)$ as
\begin{multline}
    \delta\lambda_{\varepsilon} \simeq \frac{1}{2 \sqrt{3} M (3M+r_{K, \varepsilon})}\Bigg(\frac{2( 9M^3 + 21 M^2\,r_{K, \varepsilon} - 13 M \, r_{K, \varepsilon}^2 -r_{K, \varepsilon}^3 )}{r_{K, \varepsilon}(r_{K, \varepsilon}-M)(3M+ r_{K, \varepsilon})}\delta m \\
     +  2(r_{K, \varepsilon}-5M)\delta m' + (r_{K, \varepsilon}-M) r_{K, \varepsilon}\, \delta m'' \Bigg)\Bigg|_{r=r_{K, \varepsilon}}.
\end{multline}

For $a=0$, from the expression \eqref{eq:deltaOR}, we obtain
\begin{equation}
    \delta \Omega_{\varepsilon}
    \simeq 
    \varepsilon
    \left( \frac{M}{ r_{K, \varepsilon}^{3}} \right)
    \left(
        \frac{2}{ r_{K, \varepsilon}-M}
        \sqrt{\frac{ r_{K, \varepsilon}^{3}}{M}}
    \right)
    \delta m \\ = \frac{2\, \varepsilon\sqrt{M}}{ r_{K, \varepsilon}^{3/2}( r_{K, \varepsilon}-M)}\,\delta m 
    =
    \frac{\varepsilon}{3^{3/2} M^{2}}\,\delta m ,
\end{equation}
where we have used $r_{K, \varepsilon}=3M$ for $a=0$.
By substituting $\delta m = -\frac{r}{2}\,\delta f$ into the previous result,
\begin{equation}
    \delta \Omega_{\varepsilon}
    \simeq 
    \varepsilon\,\frac{1}{3^{3/2} M^{2}}
    \left( -\frac{3M}{2}\,\delta f \right) = 
    -\varepsilon\,\frac{1}{2\sqrt{3}\,M}\,\delta f .
\end{equation}
By picking the branch $\varepsilon = -1$, or computing $\delta \hat{\Omega}_{\varepsilon}$, we recover the static result.

In a similar way, the shift of the Lyapunov exponent for $a=0$ can be evaluated as 
% , when we set $a  \xrightarrow{} 0$
\begin{equation}
\delta\lambda_{\varepsilon}
\simeq
\frac{1}{12\sqrt{3}\,M^{2}}
\left(6M^{2}\,\delta m''
-4\,\delta m
-4M\,\delta m'
\right)\Bigg|_{r=3M}.
\label{eq:dellambda_static_mid}
\end{equation}
% Then we rewrite 
Rewriting the mass perturbations using $\delta m = -\tfrac{r}{2}\,\delta f$ and substituting
the Einstein equations 
% relations 
\eqref{eq:EinsteinRel0}, we obtain
\begin{equation}
\delta\lambda_{\varepsilon}
\simeq
\left[
\frac{1}{2\sqrt{3}\,M}\,\delta f(3M)
-4\sqrt{3}\,\pi M\bigl(P_\theta(3M)-P_r(3M)\bigr)
\right].
\label{eq:dellambda_static_final}
\end{equation}
%Choosing the branch $\varepsilon=-1$, or calculating $\delta\gamma_{\varepsilon}$, 
It reproduces the static Lyapunov exponent shift.

\subsection{Modifications of QNMs and Energy Conditions}

We continue the analysis from the static case. Although an analytically simple expression for $\lambda_{\varepsilon}$ is not available, as for the static case, we can quantify the expression in terms of the hair characteristics.

 By applying the Einstein equations in the  rotating metric form \eqref{rotating:metric}, as explained in Appendix~\ref{app2}, in particular for the equatorial plane \eqref{eq:EinsteinRel0}, we can express the derivatives of $m(r)$ as 
\begin{equation}
    m'(r) = \delta m '(r) = -4 \pi r^2 P_r(r),
\label{eq:deltasRot1}
\end{equation}
\begin{equation}
    m''(r) = \delta m ''(r) = -8 \pi r P_\theta(r).
\end{equation}
We start from the Einstein equation \eqref{eq:deltasRot1} for the deviation of the mass function.
We impose the boundary condition that the deviation vanishes at infinity $ \delta m(\infty) = 0$. Integrating $\delta m'(r)$ from $r$ to $\infty$ gives
\begin{equation}
     \delta m(r) = -\int_{r}^{\infty} \delta m'(s)\, ds \,
     %= - \int_{r}^{\infty} \big( -4\pi s^{2} P_r(s) \big)\, ds \, 
     = 4\pi \int_{r}^{\infty} s^{2} P_r(s)\, ds.
     \label{eq:dmr2}
\end{equation}
For simplicity, we use \eqref{eq:dmr2} to express the components of the QNM as
 \begin{eqnarray}
\delta \hat{\Omega}_\varepsilon &=& -\varepsilon \, \delta \Omega_\varepsilon  \simeq -D_{K,\varepsilon}^{-2} \left( \frac{8 \pi}{r_{K,\varepsilon}-M}\sqrt{\frac{r_{K,\varepsilon}^3}{M}} \right) \int_{r_{K,\varepsilon}}^{\infty} s^2 P_r (s) \, ds \, ,
\label{eq:rotO2}\\
   \delta\lambda_{\varepsilon} &\simeq& 
    \,\frac{ 2 \pi}{\sqrt{3} M\,\bigl(3M + r_{K,\varepsilon}\bigr)}
    \Bigg[
        A(r_{K,\varepsilon})
        \int_{r_{K,\varepsilon}}^{\infty} s^2 P_r (s) \, ds \cr
        && \hspace{2cm}+ B(r_{K,\varepsilon})\, r_{K,\varepsilon}^2 P_r
        + 2 \, C(r_{K,\varepsilon})\, r_{K,\varepsilon} P_\theta
    \Bigg],
\end{eqnarray}
where
\begin{equation}
A(r)\equiv \frac{2\left(r^3 + 13Mr^2 - 21M^2 r - 9M^3 \right)}{r\,(r-M)\,(3M+r)}\,,
\qquad
B(r)\equiv 2(r - 5M)\,,
\qquad
C(r)\equiv (r-M)\,r\,.
\end{equation}

The sign of the shift of the UCOP radii $\delta r_\varepsilon$ \eqref{eq:deltarK} is not so clear as in the static case, while the term that contains $\delta m$ is a positive contribution, the other term is proportional to $P_r$, and there is no restriction on the sign of this term. 
The shift of the UCOP radii $\delta r_\varepsilon$ can be expressed as
 \begin{equation}
\delta r_{\varepsilon}
\simeq 
\frac{4\pi}{3M}\,r_{K,\varepsilon}^4\,P_r(r_{K,\varepsilon})
+\frac{4\pi}{3M}\frac{r_{K,\varepsilon}(r_{K,\varepsilon}+3M)}{r_{K,\varepsilon}-M}\,
\int_{r_{K,\varepsilon}}^{\infty} s^2 P_r (s) \, ds .
\label{eq:deltarr}
\end{equation}
% 
% In this scenario, t
% 
In Appendix~\ref{app2}, we derived the values for the energy-momentum tensor components at the co-rotating frame in our settings. The values for the energy density and the pressures at the co-rotating frame are the same values as in the static case. Therefore, the analysis of the energy conditions \eqref{eq:NEC}, \eqref{eq:WEC}, \eqref{eq:SEC}, and \eqref{eq:DEC} remains valid. Energy conditions are frame-independent, therefore the interpretation in the co-rotating frame should hold for any frame. Furthermore, by following the definitions~\eqref{eq:ws}, from expressions listed in Appendix~\ref{app2}, 
we can identify $w_r = -1$, and
\begin{equation}
    w_\theta = - \frac{r}{2} \frac{m''}{m'}\,.
\end{equation}
Then, according to~\eqref{eq:rotO2} and ~\eqref{eq:deltarr}, $\delta \hat{\Omega}_\varepsilon$ must be positive if we consider a positive energy density $\rho$, as seen in Appendix~\ref{app2}, and $\delta r_{\varepsilon}$, negative, as in the static case.

\section{Examples of Stationary Rotating Hairy Black Holes}
\label{ch:examplesRotating}

For rotating regular black holes, the QNM-shadow correspondence has recently been analysed in detail by Pedrotti and Vagnozzi~\cite{PedrottiVagnozzi:2024}, who verified it explicitly for rotating Bardeen and Hayward geometries under suitable separability conditions. Their results provide useful context for the rotating Bardeen and Hayward examples considered below.

\subsection{Rotating Bardeen Black Hole}

The rotating Bardeen geometry generated via the Newman-Janis prescription can be described by taking the same mass function $m(r) = M\cdot [r^2/(r^2+ q^2)]^{3/2}$ as in the static solution.
In the small-$q$ regime, one finds the following expansion
\bena
m(r) \;=\; M \Big(1 - \frac{3\,q^2}{2 r^2}\Big) \;+\; \mathcal{O}(q^4),
\qquad
m'(r) \;\simeq\; \,\frac{3 M q^2}{r^3},\qquad
m''(r) \;\simeq\; -\,\frac{9 M q^2}{r^4}.
\label{eq:rotBardeen_m_series}
\eena
On the equatorial plane ($\theta=\pi/2$), the Einstein equations for the stationary axisymmetric ansatz
\eqref{eq:rotGthth}, \eqref{eq:rotGrr}, \eqref{eq:EinsteinRel} and \eqref{eq:EinsteinRel2}
relate the derivatives of $m(r)$ to the effective anisotropic pressures. Keeping only the leading contribution in $q$, we obtain
\bena
P_r(r) \simeq -\,\frac{3 M q^2}{4\pi r^5},\qquad
P_\theta(r) \simeq \,\frac{9 M q^2}{8\pi r^5}. 
\eena

\begin{figure}[htbp]
  \centering
  \begin{subfigure}[t]{\textwidth}
    \vspace{0pt}
    \includegraphics[width=\textwidth]{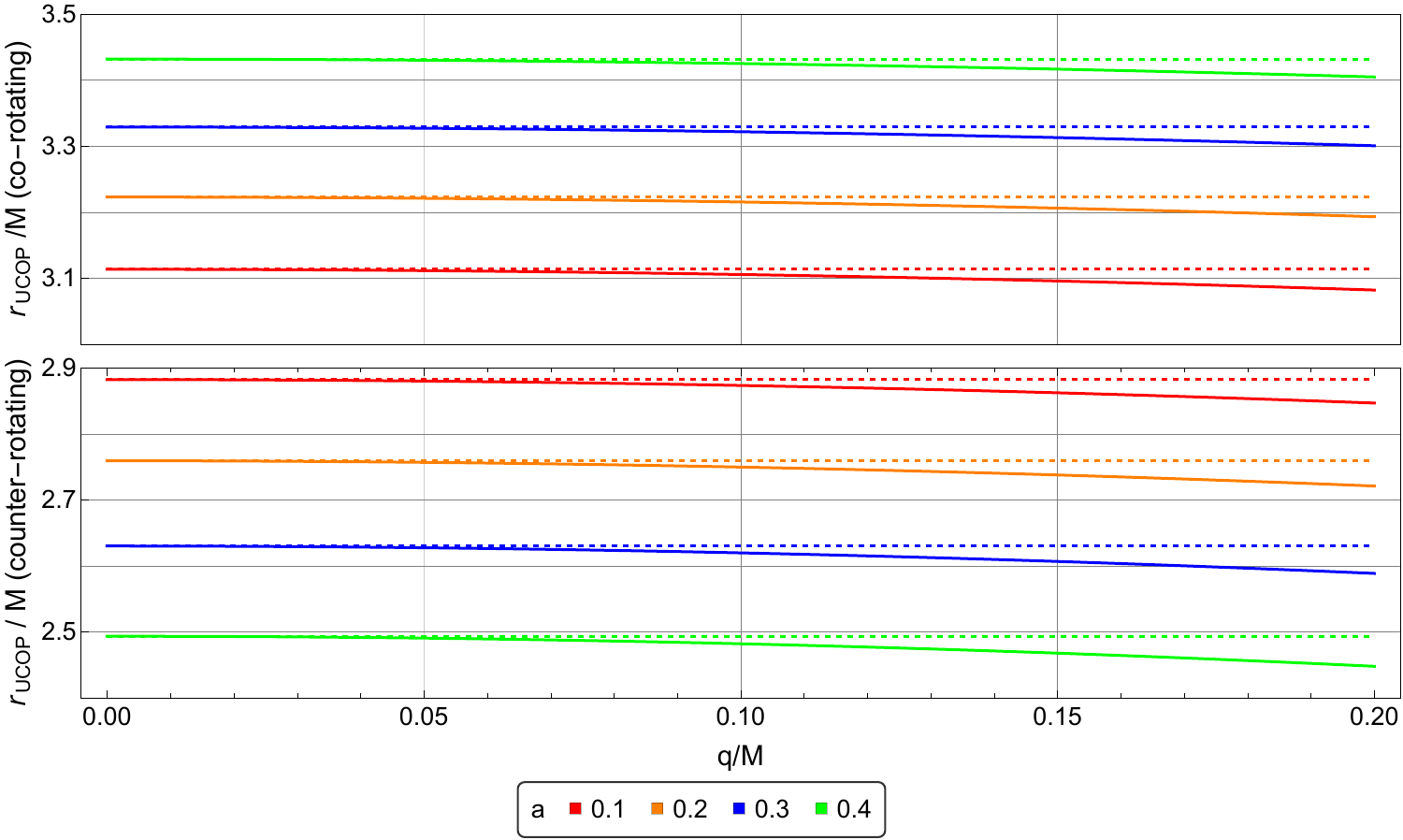}
    \caption{UCOP radius change given different $q$ parameters for different angular momentum values. The dashed lines show the cases of the Kerr solutions, and the solid lines represent the cases of the rotating Bardeen Black Hole.}
    \label{fig:fig4-1}
  \end{subfigure}
%  \hfill
 \vskip\baselineskip
  \begin{subfigure}[t]{0.45\textwidth}
   % \vspace{0pt}
    \includegraphics[width=\textwidth]{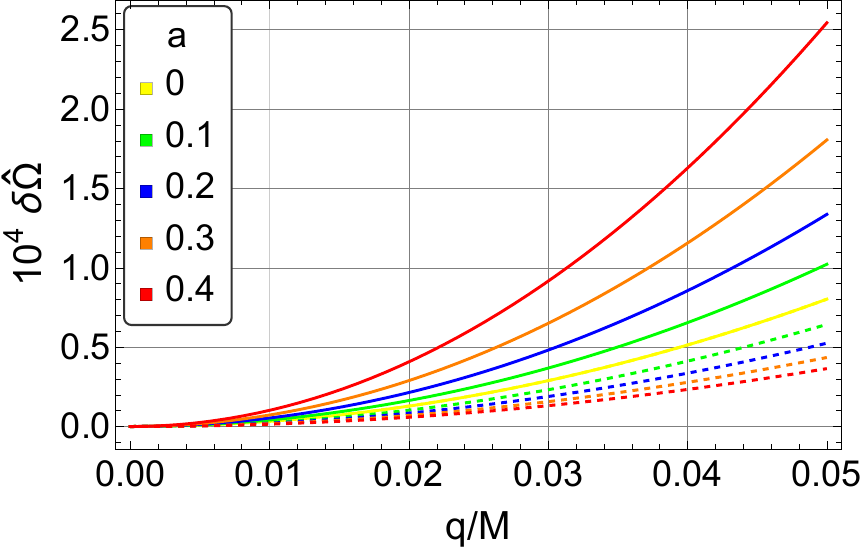}
    \caption{
    % Hair perturbation for the angular frequency. 
    Deviation of the angular frequency $\delta\hat\Omega_{\varepsilon}$ as a function of $q/M$ for several values of the spin parameter $a$.
    % Continuous 
    Solid lines and dashed lines represent the counter-rotating and co-rotating solution branches, respectively.}
    \label{fig:fig4-2}
  \end{subfigure}
%    \vskip\baselineskip
    \hfill
  \begin{subfigure}[t]{0.45\textwidth}
    \includegraphics[width=\textwidth]{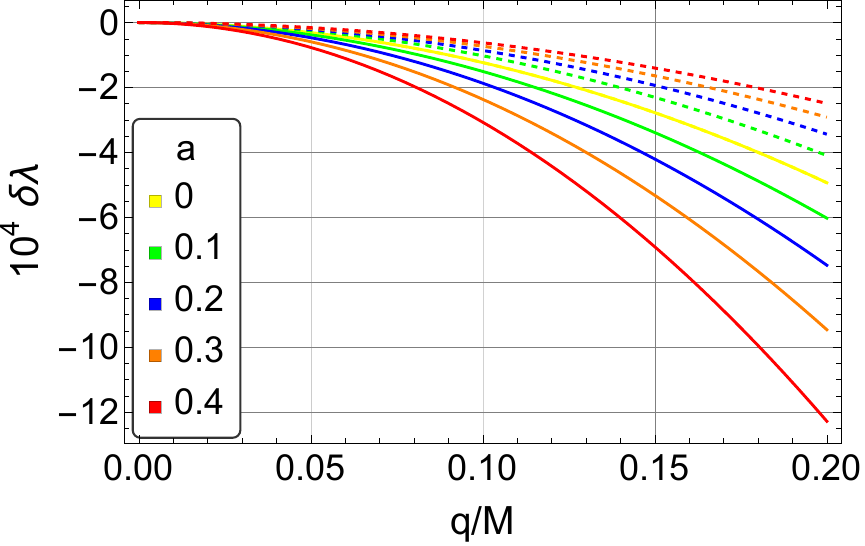}
    \caption{
    Deviation of the Lyapunov exponent $\delta\lambda_{\varepsilon}$ as a function of $q/M$ for several values of the spin parameter $a$.
    Solid lines and dashed lines represent the counter-rotating and co-rotating solution branches, respectively.}
    \label{fig:fig4-3}
  \end{subfigure}
  \caption{Leading eikonal shifts for rotating Bardeen black holes.}
  \label{fig:combinedR1}
\end{figure}

We now insert the perturbative expansion of $m(r)$ into the impact-parameter relation \eqref{eq:D1} and expand consistently to first order in the deformation parameter (i.e.\ $\mathcal{O}(q^2)$). This yields the shifts in the Kerr photon-sphere radius and in the impact parameter
\begin{align}
\delta r_{\varepsilon} 
& \simeq -\,\frac{3 r_{K, \varepsilon} + M}{2 r_{K, \varepsilon} (r_{K, \varepsilon} - M)}\,q^2 \, , 
\label{eq:eqdeltarR1}\\
\delta D_{\varepsilon}
&\simeq \frac{3 \, \varepsilon \, q^2}{r_{K, \varepsilon}-M} \sqrt{\frac{M}{r_{K, \varepsilon}}} \,.
\end{align}
We can observe in Fig.~\ref{fig:fig4-1} how the rotation splits the UCOP in two, and that as the rotation frequency $a$ increases, the separation between both radii increases. Given the presence of $q^2$ in \eqref{eq:eqdeltarR1}, we can observe that for any value of $q$, the radii of the UCOP reduce. % 

The corresponding corrections to the angular frequency and Lyapunov exponent take the forms
\bena
    \delta \hat{\Omega}_{\varepsilon} = - \varepsilon \delta \Omega_{\varepsilon} \simeq  \left( a - \varepsilon \sqrt{\frac{{r_{K, \varepsilon}}^3}{M}} \right)^{-2} \left(  \frac{3 \, q^2}{r_{K, \varepsilon}-M} \sqrt{\frac{M}{r_{K, \varepsilon}}}\right), 
\eena
\bena 
%\delta\gamma_{\varepsilon} =
%-\varepsilon 
\delta\lambda_{\varepsilon} \simeq \frac{
   2 \, \sqrt{3} M\, q^{2}
}{
    (r_{K, \varepsilon}-M)\, r_{K, \varepsilon}^{3}\, (r_{K, \varepsilon}+3M)^{2}
}
\left( 3 M^{2} - 8 M r_{K, \varepsilon} + r_{K, \varepsilon}^{2} \right). 
\eena
Fig.~\ref{fig:fig4-2} displays the deviation in the angular frequency 
$\delta\hat{\Omega}_{\varepsilon}$ of the unstable circular photon orbit as a function of the parameter $q/M$. The deviation is positive and grows monotonically with $q/M$.
The yellow line represents the case when the rotating cases reduce to the static case by setting the spin $a$ to 0. Both counter-rotating and co-rotating cases converge to a single curve, as expected. We can see that for the counter-rotating case, the magnitude of the shift produced by the hair is larger. The opposite is true for the co-rotating case.
As expected, all curves converge to $\delta\hat \Omega_{\varepsilon} = 0$ in the 
Kerr limit $q/M \to 0$.

Fig.~\ref{fig:fig4-3} shows the behaviour of the deviation parameter 
$\delta\lambda_{\varepsilon}$ as a function of 
% the hair amplitude 
$q/M$ for several values of 
the spin parameter $a$. 
The yellow line represents the static case. The magnitude of the shift for the counter-rotating case is larger in magnitude, and that of the co-rotating case is smaller.
All curves smoothly 
approach $\delta\lambda_{\varepsilon} = 0$ in the limit $q/M \to 0$, confirming 
that the Kerr solution is recovered continuously.

\subsection{Rotating Hayward Black Hole}

\begin{figure}[htbp]
  \centering
  \begin{subfigure}[t]{\textwidth}
    \vspace{0pt}
    \includegraphics[width=\textwidth]{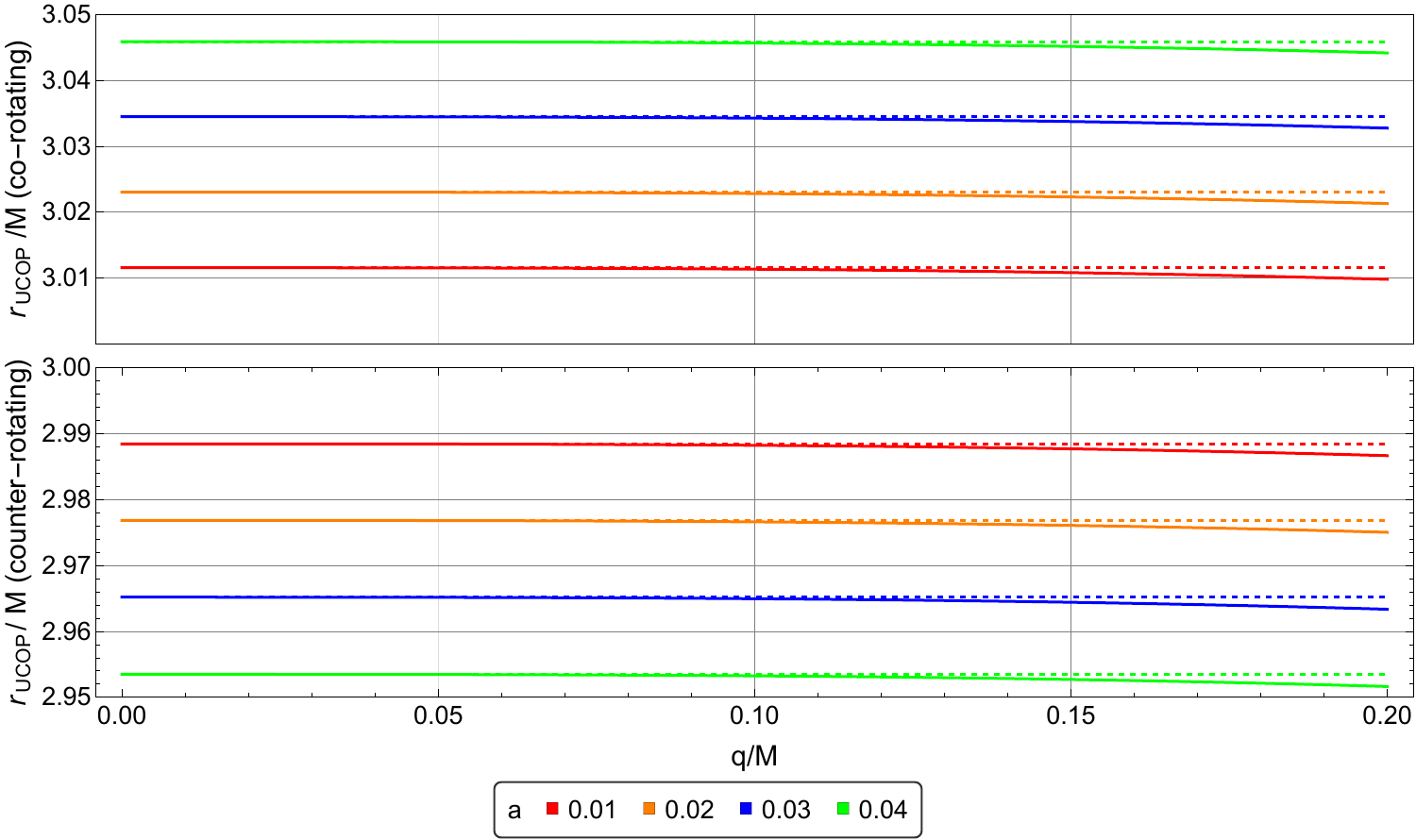}
    \caption{UCOP radius change given different $q$ parameters for different angular momentum values. The dashed lines show the cases of the Kerr solutions, and the solid lines represent the cases of the rotating Hayward Black Hole.}
    \label{fig:fig5-1}
  \end{subfigure}
    \vskip\baselineskip
  \begin{subfigure}[t]{0.45\textwidth}
    %\vspace{0pt}
    \includegraphics[width=\textwidth]{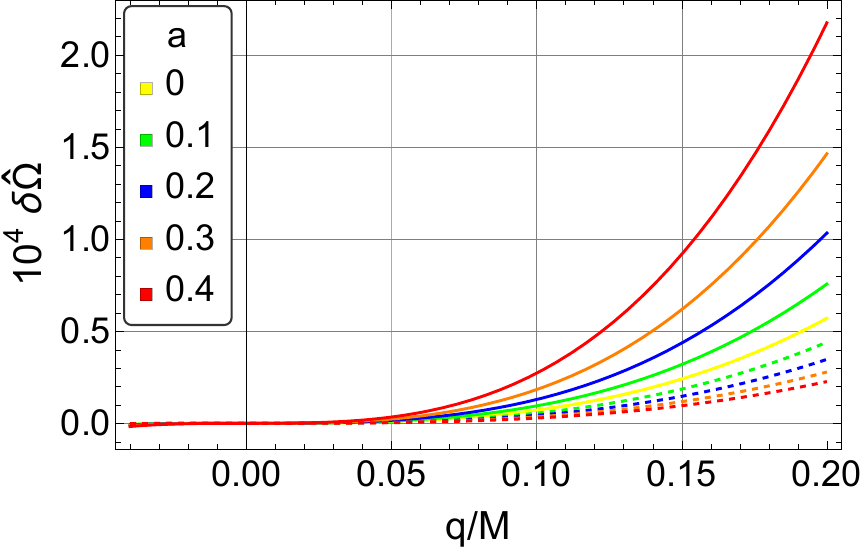}
    \caption{Deviation of the angular frequency $\delta\hat\Omega_{\varepsilon}$ as a function of $q/M$ for several values of the spin parameter $a$.}
    \label{fig:fig5-2}
  \end{subfigure}
  \hfill
  \begin{subfigure}[t]{0.45\textwidth}
    \includegraphics[width=\textwidth]{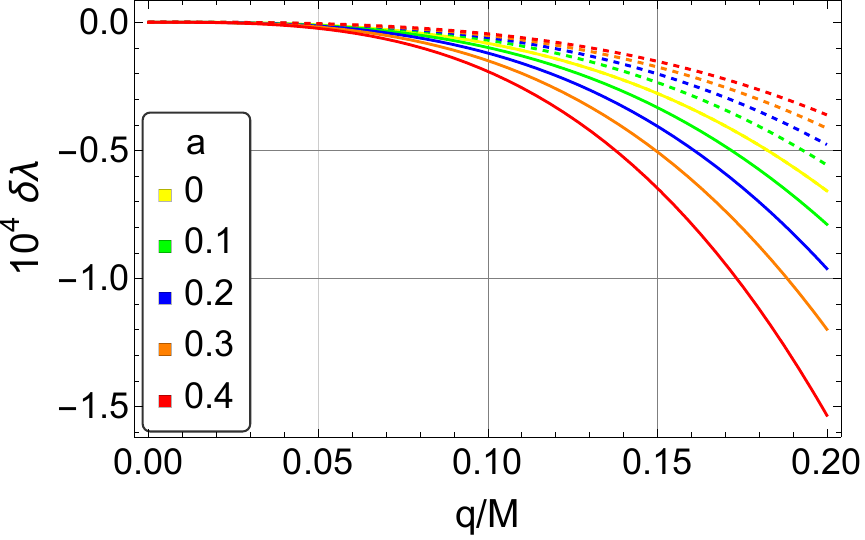}
    \caption{Deviation of the Lyapunov exponent $\delta\lambda_{\varepsilon}$ as a function of $q/M$ for several values of the spin parameter $a$.}
    \label{fig:fig5-3}
  \end{subfigure}
  \caption{Leading eikonal shifts for rotating Hayward black holes.}
  \label{fig:combinedk2}
\end{figure}

In the rotating Hayward black hole case, the mass function is defined as $m(r)=  M r^3/(r^3+ q^3)$, and for small $q$ it admits the expansion
\bena
m(r) \;=\; M \Big(1 - \frac{\,q^3}{ r^3}\Big) \;+\; \mathcal{O}(q^4),
\qquad
m'(r) \;\simeq\; \,\frac{3 M q^3}{r^4},\qquad
m''(r) \;\simeq\; -\,\frac{12 M q^3}{r^5}.
\label{eq:rotHayward_m_series}
\eena
Restricting to the equatorial plane and using the same set of field equations, we obtain, at leading order in $q$, 
\bena
P_r(r) \simeq -\,\frac{3 M q^3}{4\pi r^6},\qquad
P_\theta(r) \simeq \,\frac{3 M q^3}{2\pi r^6}. \label{eq:rotBardeen_Tmunu_firstorder}
\eena
Proceeding as in the Bardeen case, we expand \eqref{eq:D1} consistently to the first non-vanishing order. This gives
\begin{align}
\delta r_{\varepsilon} 
& \simeq -\,\frac{4 \, q^3}{3 r_{K, \varepsilon} (r_{K, \varepsilon} - M)} \, , \\
\delta D_{\varepsilon}
&\simeq  \frac{2 \, \varepsilon \, q^3}{ r_{K, \varepsilon} - M} \sqrt{\frac{M}{r_{K, \varepsilon}^3}} \, .
\end{align}
The induced corrections to the angular frequency and the Lyapunov exponent are therefore
\bena
    \delta \hat{\Omega}_{\varepsilon} \simeq  \left( a - \varepsilon \sqrt{\frac{{r_{K, \varepsilon}}^3}{M}} \right)^{-2} \left(  \frac{2 \, q^3}{r_{K, \varepsilon}-M} \sqrt{\frac{M}{{r_{K, \varepsilon}}^3}}\right), 
\eena
\bena
\delta \lambda_{\varepsilon} \simeq
\frac{
   2\, q^{3}
}{
 \sqrt{3}\,(r_{K,\varepsilon}-M)\, r_{K,\varepsilon}^{4}\,( r_{K,\varepsilon}+ 3M)^{2}
}
\left(
   9 M^{3} - 15 M^{2} r_{K,\varepsilon} - M r_{K,\varepsilon}^{2} - r_{K,\varepsilon}^{3}
\right). 
\eena
The set of plots in Fig.~\ref{fig:combinedk2} exhibits the same qualitative behaviour as the Bardeen case.  
The deviations remain smooth, monotonic as functions of the parameter $q/M$, and approach the Kerr 
limit continuously as $q/M \to 0$, with only a mild dependence on the spin parameter $a$.

\subsection{Rotating Kiselev Black Hole}

For the rotating Kiselev solution the mass function is 
\bena
m(r) = M + \frac{k}{2 r^{3 \, w_q}} \;,
\quad
m'(r) = -\,\frac{3 k\, w_q}{2 r^{3
w_q+1}}  ,\quad
m''(r) = \frac{3 k\, w_q (1+3 w_q)}{2 r^{3 w_q + 2}}  .
\label{eq:rotKiselev_m_series}
\eena
On the equatorial plane, the Einstein equations yield the effective pressures
\bena
P_r(r)= \frac{3 k\, w_q}{8\pi  r^{3(1+w_q)}},\qquad
P_\theta(r)= -\frac{3 k\, w_q\, (1 + 3 w_q)}{16\pi \, r^{3(1+w_q)}}.
\eena
Expanding the impact parameter \eqref{eq:D1} to first order in the deformation (i.e.\ linear in $k$) we obtain
\begin{align}
\delta r_{\varepsilon} 
&\simeq \frac{ k\, \left(r_{K, \varepsilon}\right)^{1 - 3 w_q}\, \Bigl( r_{K, \varepsilon} - 3M(-1 + w_q) + 3 r_{K, \varepsilon} w_q \Bigr) }{ 6M\, (r_{K, \varepsilon}-M) }\, , \\
\delta D_{\varepsilon} 
&\simeq \frac{\varepsilon \, k\, \left(r_{K, \varepsilon}\right)^{-3 w_q}\, \sqrt{(r_{K, \varepsilon})^{3}/M} }{r_{K, \varepsilon} - M} \, . 
\end{align}
Accordingly, the first-order corrections to the angular frequency and the Lyapunov exponent read
\bena
    \delta \hat{\Omega}_{\varepsilon} \simeq  \left( a - \varepsilon \sqrt{\frac{{r_{K, \varepsilon}}^3}{M}} \right)^{-2} \left(  \frac{- \, k\, (r_{K, \varepsilon})^{-3 \, w_q}\, \sqrt{(r_{K, \varepsilon})^{3}/M} }{ r_{K, \varepsilon} - M }\right), 
\eena
\begin{multline}
\delta \lambda_{\varepsilon} \simeq
\frac{
   - \, k\, r_{K, \varepsilon}^{-1 - 3 w_q}
}{
    4\sqrt{3} \, M\, (M - r_{K, \varepsilon})\, (3M + r_{K, \varepsilon})^{2}
}
\left(
    9 M^{3}\!\left(2 + 3(-3 + w_q)w_q\right) \right. \\
    \left.+ 3 M^{2}\!\left(14 + 3(7 - 5w_q)w_q\right) r_{K, \varepsilon}  
    + M\!\left(-26 + 3w_q(7 + 3w_q)\right) r_{K, \varepsilon}^{2}\right.
    \\
    \left. + (-2 + 3w_q)(1 + 3w_q)\, r_{K, \varepsilon}^{3}
\right). 
\end{multline}

\begin{figure}[htbp]
  \centering
  \begin{subfigure}[t]{\textwidth}
    \vspace{0pt}
    \includegraphics[width=\textwidth]{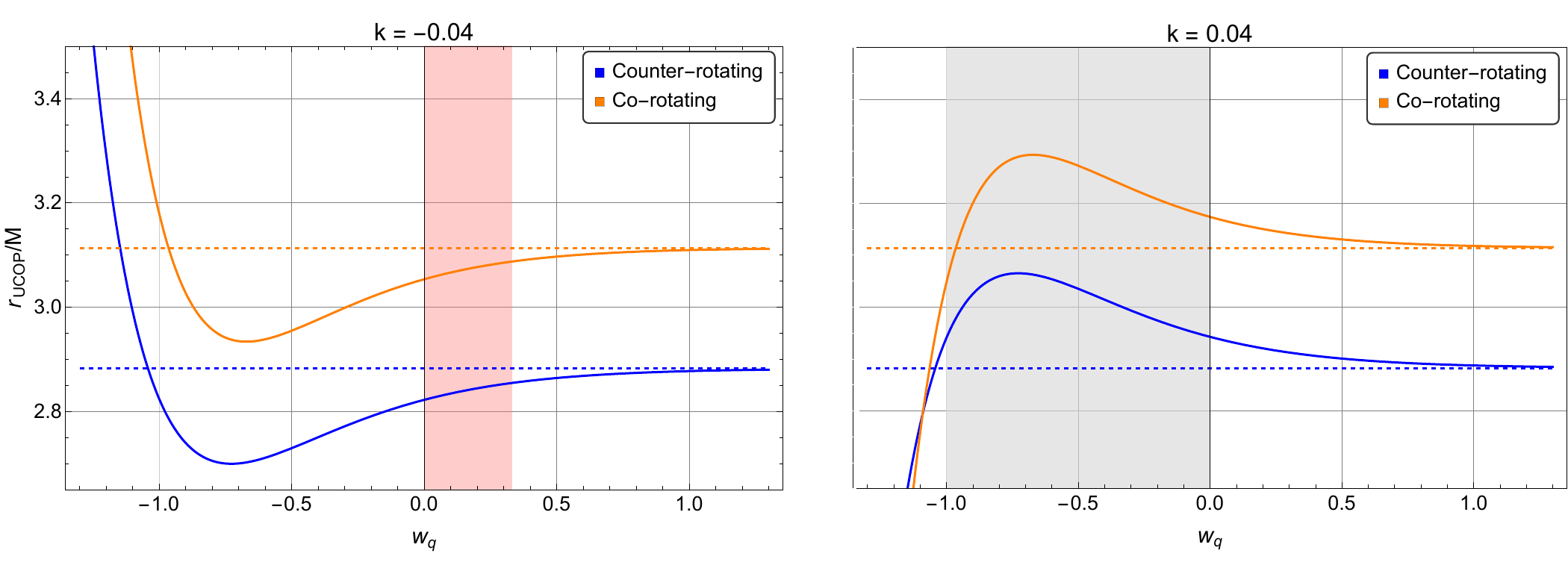}
    \caption{UCOP radius change given different $w_q$ parameters. Blue represents the counter-rotating branch solution and orange the co-rotating branch solution. The dotted lines show the cases of the Kerr solutions, and the dashed lines represent the cases of the rotating Kiselev black hole model. The shaded areas denote the branch that fulfils the DEC. Accordingly, for $k>0$, the grey interval satisfies the DEC, whereas for $k<0$, the pink interval does.}
    \label{fig:fig6-1}
  \end{subfigure}
  \begin{subfigure}[t]{\textwidth}
     \includegraphics[width=0.32\textwidth]{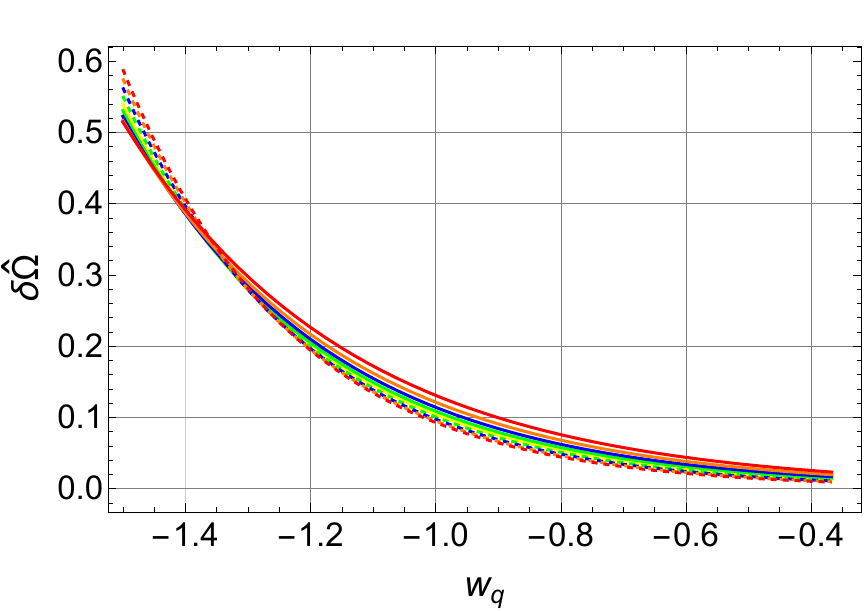}
    \includegraphics[width=0.32\textwidth]{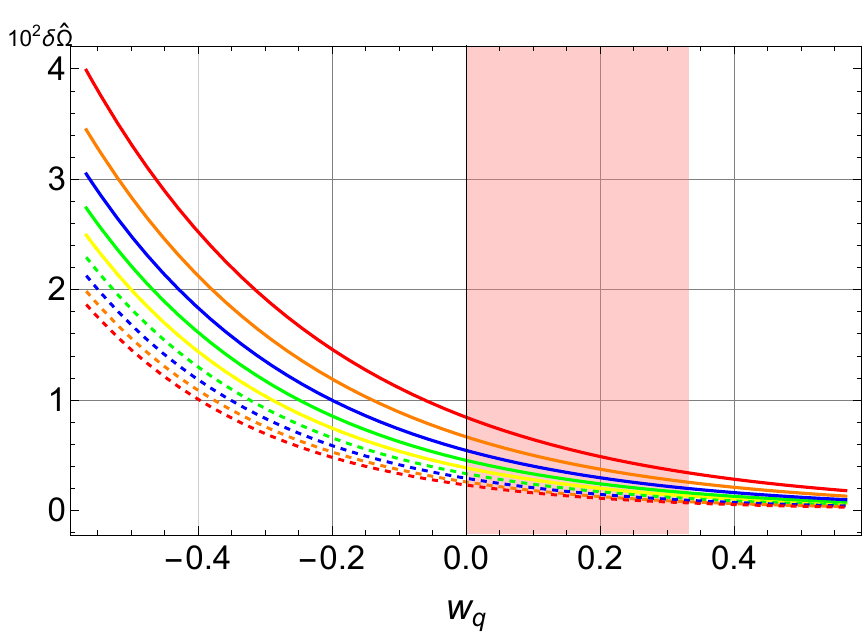}
    \includegraphics[width=0.32\textwidth]{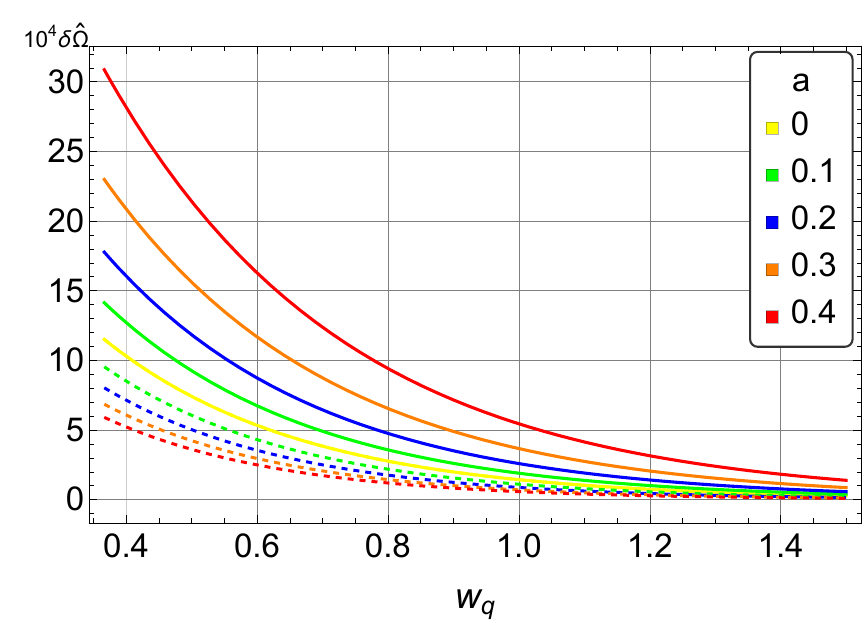}
    \caption{Deviation of the angular frequency $\delta\hat\Omega_{\varepsilon}$ as a function of $w_q$ for several values of the spin parameter $a$ with $k=-0.04$. The shaded area represents the interval that fulfils the DEC.}
    \label{fig:fig6-2}
\end{subfigure}
    \vskip\baselineskip
  \begin{subfigure}[t]{\textwidth}
    \includegraphics[width=0.32\textwidth]{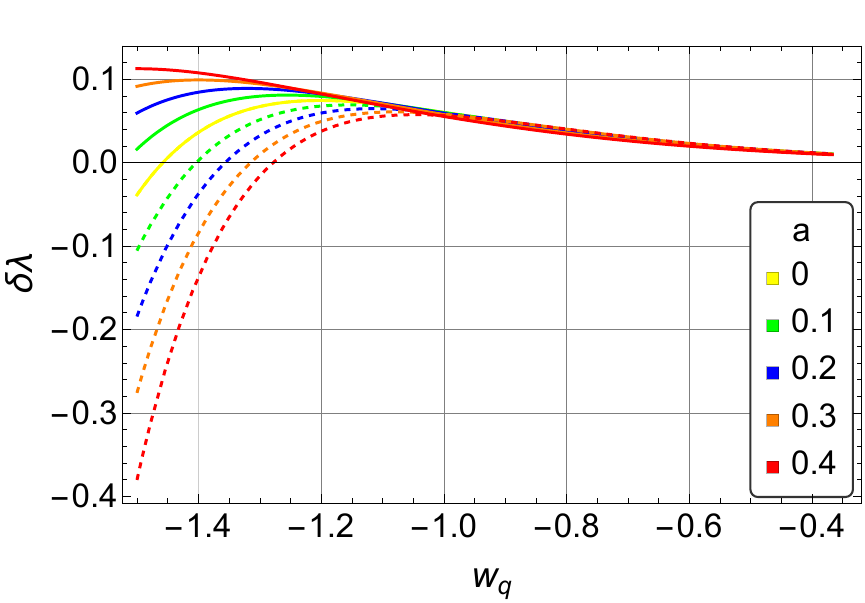}
    \includegraphics[width=0.32\textwidth]{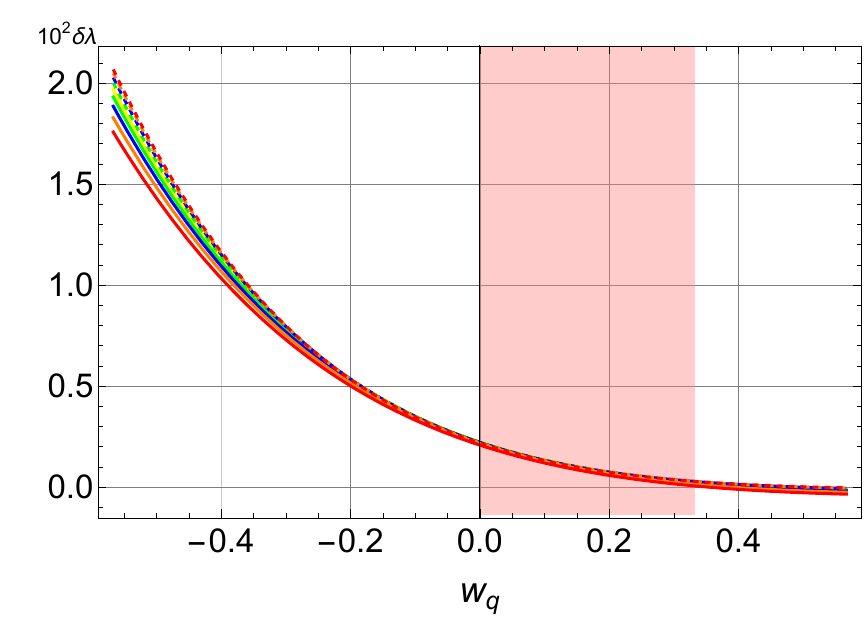}
    \includegraphics[width=0.32\textwidth]{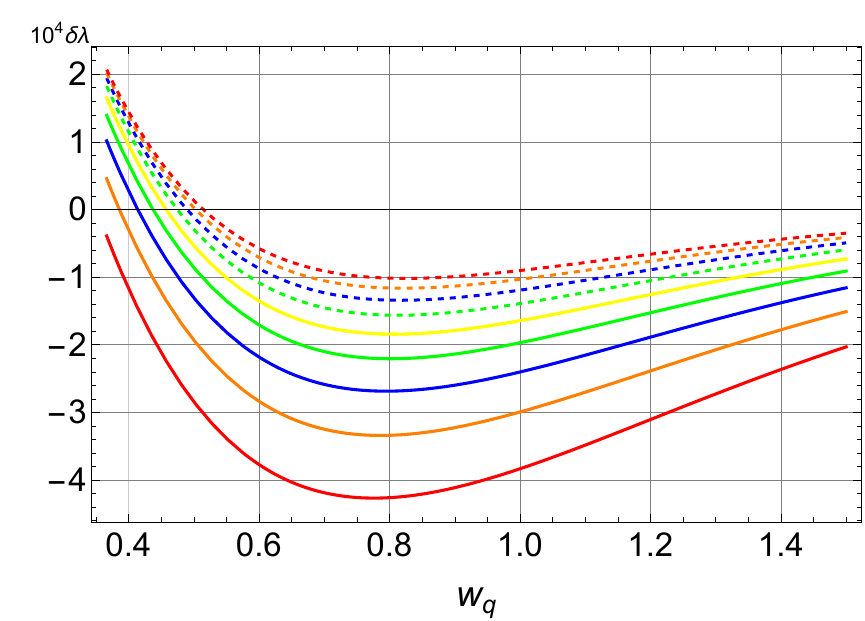}
    \caption{Deviation of the Lyapunov exponent $\delta\lambda_{\varepsilon}$ as a function of $w_q$ for several values of the spin parameter $a$ with $k=-0.04$. The shaded area represents the interval that fulfils the DEC.}
    \label{fig:fig6-3}
    \end{subfigure}
  \caption{Leading eikonal shifts for rotating Kiselev black holes.}
  \label{fig:combinedk1}
\end{figure}
%\FloatBarrier

The behaviour of the QNM coefficients for the Kiselev black hole, shown in
Figs.~\ref{fig:combinedk1}, can be interpreted 
directly from the effective matter content of the model. The parameters $k$ and $w_q$
govern the effective energy-momentum distribution, which in turn determines
the sign and radial profile of the deviation $\delta m(r)$ and the anisotropic
pressures $(P_r,P_\theta)$. 

When $w_q$ is in the range of quintessence, the shift for the Lyapunov exponent is positive. Unlike the Bardeen and Hayward cases, where the matter distribution tended to reduce the instability, here the effective matter contribution enhances the divergence rate of nearby null geodesics. Outside of this range, we can find values of $w_q$ that produce the same sign values of $\delta \Omega$ and $\delta \lambda$ as in the previous cases.
This is clearly visible in the Figs.~\ref{fig:fig6-2} and \ref{fig:fig6-3}.

\section{Summary and discussion}\label{ch:summary}

In this work, we developed a unified perturbative framework to study QNM frequencies of stationary hairy black holes by exploiting the correspondence between the unstable circular null geodesics and QNMs.  
We modelled deviations from the corresponding vacuum solutions of the Einstein equations as perturbations in the form of an anisotropic fluid.
Using the relation between the eikonal limit of QNMs and unstable null circular orbits, we derived explicit formulas for the QNM frequencies in terms of the state parameters of the hair field.
Our formulas provide a systematic approach to the analysis of QNMs for hairy black holes without assuming any specific underlying theory or model. We also examined the relationship between the state parameters and the QNM frequencies, including possible violations of the energy conditions. 

In the static case, we showed that the QNM shifts can be expressed directly in terms of the matter distribution and its equation of state, with the Lyapunov exponent receiving an additional explicit contribution from the tangential pressure.
In particular, for the Bardeen and Hayward black hole models, the small-deviation approximation from the vacuum solution is incompatible with the dominant energy condition (DEC).
This implies that, if the corresponding deviation is confirmed through the QNM observations, we need to consider a matter field violating the DEC or any other modification in the gravitational wave emission mechanisms. 
Since the violation of the DEC implies a superluminal flux, it might be more physically reasonable to consider a modified gravity theory than to rely on exotic matter fields. By contrast, for the Kiselev case, the compatibility with the energy conditions depends on the parameters $k$ and $w_q$, and all of the standard energy conditions can be satisfied within suitable parameter ranges even in the small-$|k|$ regime. For all the physical examples studied in this work, the WEC and SEC are compatible with the perturbative regimes considered. Thus, the effective source can preserve positive energy density and satisfy the usual attractive-energy requirements associated with it.
   
In the rotating case, the QNM components are separated into co-rotating and counter-rotating branches according to the equatorial UCOP structure. This shows that rotation does not merely shift the static result, but introduces an asymmetry in how the hair modifies the oscillation frequency and damping rate, with both branches reducing smoothly to the static case in the non-rotating limit. We considered the Bardeen, Hayward, and Kiselev metrics obtained by applying the Newman--Janis method to their static counterparts.
We then restricted our attention to UCOPs on the equatorial plane and derived the corresponding angular frequencies and Lyapunov exponents, from which the QNM information can be inferred. Moreover, because the effective energy density and principal pressures in the co-rotating frame are the same as in the corresponding static case, the energy-condition analysis performed for the static models remains applicable to the rotating case as well.
% Although the correspondence between QNMs and UCOPs is not fully established in the rotating case, equatorial UCOPs are expected to be associated with the $\ell=|m|$ mode of QNMs \cite{Cardoso:2008bp}.
In the rotating case, the correspondence between QNMs and UCOPs is more subtle than in the static spherically symmetric case because the azimuthal degeneracy is
broken. In this work, we restrict our analysis to equatorial unstable circular photon orbits, which are associated with the eikonal $\ell=|m|$ sector in the established literature \cite{Dolan:2010wr,Yang:2012he,Jusufi:2022}. Therefore, the rotating formulas derived here should be interpreted as leading-eikonal results for the co-rotating and counter-rotating equatorial branches, not as a calculation of the full rotating QNM spectrum.

The connection with gravitational-wave observables is then direct at the level of leading eikonal shifts. The real part of the QNM frequency determines the ringdown frequency, $f_{\rm RD}=\mathrm{Re}(\omega)/(2\pi M_z)$, while the imaginary part determines the damping time, $\tau_{\rm RD}=M_z/|\mathrm{Im}(\omega)|$, where $M_z$ is the redshifted remnant mass~\cite{Berti:2009kk}. Therefore, within the eikonal approximation, $\delta\Omega/\Omega$ approximately gives the fractional shift of the ringdown frequency, while $-\delta\lambda/\lambda$ approximately gives the fractional shift of the damping time. For observed ringdown modes, especially the dominant $(\ell,\mathbf{m},n)=(2,2,0)$ mode, a quantitative comparison would require a dedicated perturbation equation for each hairy metric, calibration against numerical QNM calculations, or a parametrized ringdown treatment~\cite{Konoplya:2022gjp,Giesler:2019uxc}. Such dedicated wave analyses are beyond the scope of the present work, but they constitute a natural next step. The leading-eikonal formulas derived here can provide analytic guidance for such studies by identifying which effective matter variables control the frequency and damping shifts.

In the construction of our hairy black hole models, we considered the effects of matter fields as perturbative deviations from the Schwarzschild and the Kerr solutions. In particular, we assumed the simplest form of the equations of state~\eqref{eq:ws} and focused only on the first-order perturbative effects of the matter field. It should be noted that our formulas given in Sec.~\ref{subsec:static:formulas} (also the formula~\eqref{eq:lambda:gen}) for the static case, and those in Sec.~\ref{subsec:rotating:formulas} (see e.g., \eqref{eq:OmegaR0}, \eqref{eq:lamdaR0}) for the rotating case can be applied to more general cases. It would therefore be interesting to extend our present analyses to higher-order perturbative analyses, just like higher-order WKB analyses, and to include more general equations of state, such as the polytropic models.

Overall, our results suggest that black hole ringdown signals can probe not only the spacetime geometry, but also the physical nature of the matter environment and possible deviations from vacuum general relativity in the vicinity of black holes.
\bigskip
\goodbreak
\centerline{\bf Acknowledgments}

\medskip 
\noindent
This work was supported in part by JSPS KAKENHI Grant No.~JP24K07027 (C.Y.), JP25K07281 (C.Y.), JP25K07306 (A.I.), and also supported by MEXT KAKENHI Grant-in-Aid for Transformative Research Areas A Extreme Universe No.~JP21H05182(A.I.) and JP21H05186(A.I.). 

\appendix

\section{Einstein tensor for static spherically symmetric metric} \label{app1}

Let us consider a spherically symmetric, static metric of the form 
\bena
 ds^2 = - f(r) dt^2 + h(r) dr^2 + r^2 (d\theta^2 + \sin^2 \theta d \varphi^2) 
 \,. 
 \label{ansatz:stat:metric}
\eena
The non-vanishing components of the connection coefficients are 
\bena {\mit \Gamma}^t_{tr} = \dfrac{f'}{2f} \,, \quad  
{\mit \Gamma}^r_{tt}=\dfrac{f'}{2h} \,, \quad {\mit \Gamma}^r_{rr} = \dfrac{h'}{2h} \,, \quad {\mit \Gamma}^r_{\theta \theta} = - \dfrac{r}{h} \,,  \quad {\mit \Gamma}^r_{\varphi \varphi}  = {\mit \Gamma}^r_{\theta \theta} \sin^2 \theta \,, \quad {\mit \Gamma}^\theta_{\theta r}= {\mit \Gamma}^\varphi_{\varphi r} = \dfrac{1}{r} \,.
\eena
The Einstein tensor is given by the components 
\bena
 G^t{}_t &=& -\dfrac{1}{rh}\left[ \dfrac{h'}{h} + \dfrac{1}{r}(h-1) \right] \,, 
 \\
 G^r{}_r &=& -\dfrac{1}{rh}\left[ -\dfrac{f'}{f} + \dfrac{1}{r}(h-1) \right] \,, 
 \\
 G^\theta{}_\theta = G^\varphi{}_\varphi &=& \dfrac{1}{2\sqrt{fh}}\dfrac{d}{dr}\left(\dfrac{f'}{\sqrt{fh}} \right)   + \dfrac{1}{2rh} \left(\dfrac{f'}{f} - \dfrac{h'}{h} \right) 
 \nonumber \\
  &=& \dfrac{1}{rh} \left[ \dfrac{r}{2} \dfrac{f''}{f} - \dfrac{r}{4} \left(\dfrac{f'}{f}\right)^2  - \dfrac{4}{r}\dfrac{f'}{f}\dfrac{h'}{h} +\dfrac{1}{2} \left(\dfrac{f'}{f} - \dfrac{h'}{h}\right) \right] \,,  
\\
G^\theta{}_\theta &=& \frac{r}{2} \dfrac{d G^r{}_r}{dr} + \frac{1}{2}(G^t{}_t + G^r{}_r). 
\eena 

\section{Einstein tensor for a class of stationary axisymmetric metric} \label{app2}

Let us consider the following form of a stationary axisymmetric metric~(\ref{rotating:metric})
\bena
ds^2&=& -f dt^2-2a\sin^2\theta (1- f)dtd\varphi + \left\{\Sigma +(2-f)a^2\sin^2\theta \right\} \sin^2\theta d\varphi^2 
\non \\
 &{}& \qquad 
+ \dfrac{\Sigma}{\Sigma f +a^2 \sin^2\theta} dr^2 + \Sigma d\theta^2 \,,
\non
\\
 &{}& f :=1-\dfrac{2m(r)r}{\Sigma} \,, \quad \Sigma :=r^2+a^2 \cos^2\theta \,, 
\label{ansatz:rotating:metric}
\eena
with $m(r)$ being an arbitrary differentiable function of $r$ and $a$ the spin parameter. 
% \medskip 
The Einstein tensor for the metric~(\ref{ansatz:rotating:metric}) is given by the components
\bena
 G^t{}_t &=& \dfrac{1}{\Sigma^3}\left\{ a^2 r \sin^2 \theta \Sigma m'' + 2m'[ a^4 \cos^2\theta \sin^2\theta -a^2r^2 -r^4] \right\} \,, 
 \\
 G^t{}_\varphi &=& - \dfrac{1}{\Sigma^3}a \sin^2 \theta \left\{ r(a^2+r^2) \Sigma m'' + 2m'[ a^4 \cos^2\theta -a^2r^2 \sin^2\theta  -r^4] \right\} \,, 
 \\
 G^\varphi{}_\varphi &=& - \dfrac{1}{\Sigma^3} \left\{ r(a^2+r^2) \Sigma m'' + 2a^2m'[ \Sigma -2r^2 \sin^2\theta] \right\} \,, 
 \\
 G^\theta{}_\theta &=& - \frac{1}{\Sigma^2} \left\{ r \Sigma m'' + 2a^2 \cos^2\theta m' \right\} \,,  
\label{eq:rotGthth}
 \\
 G^r{}_r &=& -\dfrac{2r^2m'}{\Sigma^2} \,. 
\label{eq:rotGrr}
\eena
% \medskip 
For the case of $\theta=\pi/2$, we find the following relations 
\bena
 G^t{}_t &=& G^r{}_r + \dfrac{a^2}{r^2}(G^r{}_r -G^\theta{}_\theta ) \,, 
 \\
 G^t{}_\varphi &=& - a \left( 1+\dfrac{a^2}{r^2} \right) (G^r{}_r -G^\theta{}_\theta ) \,, 
 \\
 G^\varphi{}_\varphi &=& G^\theta{}_\theta - \dfrac{a^2}{r^2} (G^r{}_r -G^\theta{}_\theta ) \,, 
 \\
 G^{\varphi}{}_t &=&  \frac{a}{r^2} (G^r{}_r -G^\theta{}_\theta ) \,. 
 \label{eq:EinsteinRel0}
\eena

The non-trivial components of the Einstein equations take the form
\bena
 %G^t{}_t= - 8\pi \rho \,, \quad 
 G^r{}_r= 8\pi P_r \,, \quad 
 G^\theta{}_\theta = 
 %G^\varphi{}_\varphi = 
 8\pi P_\theta \,.  
 \label{eq:EinsteinRel}
\eena
From these, we find 
\bena
 m = M-4\pi \int dr r^2 P_r \, , \quad \,%, \quad - \rho=P_r  + \dfrac{a^2}{r^2}(P_r - P_\theta) \,. 
 P_r = - \frac{m'}{4 \pi r^2}\,,\quad
 P_{\theta} = - \frac{m''}{8 \pi r}\,.
  \label{eq:EinsteinRel2}
\eena

Because of the model rotation, $G^t_t$ no longer resembles a physical density. In particular, a non-vanishing cross component $T^{t}_\varphi =  G^t_{\varphi}/(8\pi)$ signals the presence of an azimuthal energy flux, so that $T^t{}_t$ and $T^\varphi{}_\varphi$ are no longer the density and the pressure measured in the fluid co-moving frame. 
To achieve a physically meaningful value, we therefore should take the co-moving frame in which the $\{t,\varphi\}$ block of the mixed tensor is diagonal. 
To do this, we diagonalize the sub-matrix of $T^\mu{}_\nu$ restricted to the $\{t,\varphi\}$ subspace. Its eigenvalues are invariant and yield $T'^{\mu}_{\nu}=\mathrm{diag}(-\rho,\,P_\varphi)$, where the identification relies on its associated eigenvector. The eigenvector associated to $\rho$ must be timelike, which defines the co-moving frame and the one associated to $P_\varphi$, spacelike, which defines the principal azimuthal direction in that frame.  
The values of $\rho$ and $P_\varphi$ in the co-moving frame therefore become

\begin{equation}
    \rho = \frac{m'}{4 \pi r^2}\,,
\end{equation}
\begin{equation}
    P_{\varphi} = - \frac{m''}{8 \pi r}\,.
\end{equation}
Since the observer in the co-moving frame can be defined as $u^\nu = (1,0,0,0)$, $\rho
=
\frac{1}{8\pi}G_{\mu\nu}u^\mu u^\nu
$. 

\section{Generalized Tolman–Oppenheimer–Volkoff Equation from Einstein’s Field Equations for Anisotropic Fluids}\label{app4}

Let us consider a general spherically symmetric static metric as given in \eqref{ansatz:stat:metric}, along with a stress-energy tensor $T^\mu_\nu$ of the form \eqref{eq:stressenergy}.

By solving Einstein's equations for the temporal and radial components, we obtain the following expressions
\bena
    h(r)=\left(1-\frac{2m(r)}{r}\right)^{-1} ,\quad  f'(r) = \frac{f(r)}{r} \left(h(r)(8\pi r^2 P_r +1) - 1 \right). 
\label{eq:tov1}
\eena
Combining these results yields an explicit expression for $f'(r)$. 
We obtain an equation from the radial component of the conservation of the stress-energy tensor $\nabla_\mu {T^\mu}_r=0$ as
\bena
    \frac{dP_r}{dr} = - \frac{f'(r)}{2 f(r)}(P_r + \rho) + \frac{2}{r} (P_\theta - P_r).
\label{eq:tov2}
\eena
Finally, substituting \eqref{eq:tov1} into \eqref{eq:tov2}, we obtain an expression that relates tangential pressure $P_\theta$ and radial pressure $P_r$
\bena
    P_\theta(r) = P_r(r) + \frac{r}{2} P'_r(r) + \frac{4 \pi r^3 P_r(r) +m(r)}{2(r - 2m(r))}(P_r(r) + \rho).
\eena

\end{document}